\documentclass[manuscript]{aastex}
\usepackage{subfigure}
\usepackage{natbib}
\usepackage{url}
\newcommand{\msun}{\mbox{$M_\odot$}}

\newcommand{\hst}{\emph{HST}}
\newcommand{\chandra}{\emph{Chandra}}
\newcommand{\ergs}{\mbox{${\rm erg}~{\rm s}^{-1}$}}
\newcommand{\ergsc}{\mbox{${\rm erg}~{\rm s}^{-1}~{\rm cm}^{-2}$}}
\newcommand{\bi}{\mbox{$B\!-\!I$}}

\shortauthors{Maxwell et al.}
\shorttitle{X-ray Binaries in NGC~6388}

\begin{document}

\title{
X-ray Binaries in the Ultrahigh
  Encounter Rate Globular Cluster NGC~6388}

%\author{J. Edward Maxwell, Phyllis M. Lugger, \& Haldan N. Cohn} 
%\author{Indiana Group}
%\affil}{Department of Astronomy, Indiana University, 727 E. Third St.,
%Bloomington, IN 47405; tmaxwell@astro.indiana.edu, lugger@astro.indiana.edu,
%cohn@astro.indiana.edu }

%\author{Jonathan E. Grindlay} 
%\affil{Harvard College Observatory, 60 Garden St., MS-6, Cambridge, MA
%02138; josh@cfa.harvard.edu}

%\author{Craig O. Heinke}
%\affil{Ingenuity New Faculty, Department of Physics, University of Alberta, Edmonton, AB T6G 2G7, Canada; cheinke@ualberta.ca}

%\author{Sonia A. Budac}
%\affil{Department of Physics, University of Alberta, Edmonton, AB T6G 2G7, Canada}

%\author{Gordon A. Drukier}
%\affil{Advanced Fuel Research, Inc., 87 Church Street, East Hartford, CT 06108}

%\and

%\author{Charles D. Bailyn}
%\affil{Department of Astronomy, Yale University, P.O. Box 208101 New Haven, CT 06520}

\author{J. Edward Maxwell\altaffilmark{1},
Phyllis M. Lugger\altaffilmark{1}, 
Haldan N. Cohn\altaffilmark{1}, 
Craig O. Heinke\altaffilmark{2},
Jonathan E. Grindlay\altaffilmark{3}, 
Sonia A. Budac\altaffilmark{4},
Gordon A. Drukier\altaffilmark{5},
Charles D. Bailyn\altaffilmark{6}}
%\author{Indiana Group}
\altaffiltext{1}{Department of Astronomy, Indiana University, 727 E. Third St.,
Bloomington, IN 47405; tmaxwell@astro.indiana.edu, lugger@astro.indiana.edu,
cohn@astro.indiana.edu }
\altaffiltext{2}{Ingenuity New Faculty, Department of Physics, University of Alberta, Edmonton, AB T6G
2G7, Canada; heinke@ualberta.ca}
\altaffiltext{3}{Harvard College Observatory, 60 Garden St., MS-6, Cambridge, MA
02138; josh@cfa.harvard.edu}

\altaffiltext{4}{Department of Physics, University of Alberta, Edmonton, AB T6G 2G7, Canada}

\altaffiltext{5}{Advanced Fuel Research, Inc., 87 Church Street, East Hartford, CT 06108}

\altaffiltext{6}{Department of Astronomy, Yale University, P.O. Box 208101
   New Haven, CT 06520}

   \begin{abstract}

We report the results of a joint \chandra-\hst\ study of the X-ray binary
population in the massive, high-density globular cluster NGC~6388.  NGC~6388
has one of the highest predicted X-ray binary production rate of any Galactic
cluster.  We detected a large population of 61 \chandra\ sources within the
half-mass radius with L$_X > 5 \times 10^{30}$ \ergs.  From the X-ray colors,
luminosities, (lack of) variability, and spectral fitting, we identify five as
likely quiescent low-mass X-ray binaries.  Due to the extremely crowded nature
of the core of NGC~6388, finding optical identifications to \chandra\ sources
is challenging. We have identified four blue, optically variable counterparts
to spectrally hard X-ray sources, evidence that these are bright cataclysmic
variables (CVs).  One showed variability of 2 magnitudes in V, indicative of a
dwarf nova eruption.  One other likely CV is identified by its X-ray spectrum
(partial covering with high $N_H$) and strong variability, making five likely
CVs identified in this cluster.  The relatively bright optical magnitudes of
these sources put them in the same class as CV1 in M15 and the brightest CVs in
47 Tuc.

\end{abstract}  


\section{Background}

Studying the relationship between X-ray binaries (XRBs) and globular clusters
continues to provide interesting clues into cluster dynamical evolution. 
These clusters have long been thought to be major producers of XRBs, since
dense cluster cores promote the type of dynamical interactions that lead to
tight binaries such as XRBs \citep{Clark75,Pooley03}.  These interactions
include exchange interactions where a massive stellar remnant such as a heavy
white dwarf or neutron star has a close encounter with an existing binary,
replacing one of the other stars \citep{Hills76}, tidal capture by a heavy
remnant of a main-sequence star \citep{Fabian75}, or collisions of neutron
stars with red giants, eventually producing ultracompact binaries with neutron
stars accreting from white dwarfs \citep{Verbunt87}.  Cataclysmic variables
(CVs) and low-mass X-ray binary (LMXB) systems are common products of these
interactions \citep{Ivanova06,Ivanova08}.  In globular clusters, a significant
number of millisecond pulsars (MSPs) have been identified, such as in 47 Tuc
\citep{Camilo00}.  Such systems are the likely progeny of LXMBs
\citep{Archibald09}. Since mass segregation causes the more massive populations
of binary stars and degenerate remnants (for instance, MSPs) to be more
centrally concentrated, these objects will be strongly tied to the dynamical
evolution of the cluster. Most significantly, this increases the likelihood of
further interactions of binaries in the core which can then halt the progress
of dynamical relaxation, supporting the core against core collapse
\citep{Fregeau08}.

The known distances, reddenings, and ages of globular clusters allow for an
analysis of populations of CVs and LMXBs that is not possible for field
objects. Using the Hubble Space Telescope (\hst) in conjunction with the
Chandra X-ray Observatory (\chandra)  large numbers of X-ray sources have been
identified in clusters and matched to optical counterparts. For instance,
\citet{Heinke05} detected 300 X-ray sources down to $L_X \sim 8\times10^{29}$
\ergs~ within the half-mass radius of 47 Tuc, 105 of which could be securely
classified as quiescent LMXBs (qLMXBs), CVs, MSPs, or chromospherically active
binaries (ABs), mostly by using HST to identify and characterize optical
counterparts \citep{Edmonds03,Edmonds03b}.  Similar identifications were also
made of 59 of the 79 X-ray sources detected (with limiting $L_X \sim 10^{29}$
\ergs) in NGC~6397, a cluster that has undergone core-collapse, by
\citet{Cohn10} in a joint \hst-\chandra\ study.

Higher numbers of interactions in cluster cores lead to larger populations of
X-ray sources \citep{Pooley03}.  This interaction rate can be simply estimated
by the encounter rate, $\Gamma \propto \rho_0^2 r_c^3/v_0$
\citep{Verbunt87b,Maccarone11}.  This quantity is a measure of how often
stellar interactions will occur as a function of the cluster's central density,
core radius and central velocity dispersion. The numbers of X-ray sources in a
cluster, particularly the qLMXBs, MSPs, and brighter CVs, vary directly with
this encounter rate, implying strongly that dynamical interactions are the
primary source of these populations in globular clusters
\citep{Pooley03,Heinke03,Pooley06}, while fainter ABs seem to be largely
primordial in origin \citep{Bassa04}.

\subsection{NGC~6388\label{ngc6388}}

NGC 6388 is a particularly good candidate for the study of the dynamical origin
of XRBs in dense-core globular clusters due to its large central density
($10^{5.34}~L_\odot{\rm pc}^{-3}$) and core radius ($7.2'' = 0.35~{\rm pc}$ at
9.9 pc from the Sun)
\citep{Harris96}\footnote{\url{http://physwww.physics.mcmaster.ca/~harris/mwgc.dat}
for 2010 revision.}. These give it one of the largest values of $\Gamma$ of any
globular cluster in the Galaxy. With such a large encounter rate, we expect it
to host a rich population of LMXBs and CVs, and thus to find a large number of
X-ray sources. In \citet{Lugger87}, the U-band surface brightness profile,
which is dominated by horizontal branch stars,  was found to be well fit by a
single mass King model, suggesting that the core is likely being supported by a
large number of primordial binaries. It is possible that some of these
primordial binaries could be detected now as ABs.

NGC 6388 also has several features beyond its large encounter rate that make it
a cluster worth investigating. For instance, for a cluster of its metallicity
(as high as ${\rm [Fe/H]} \approx -0.7$ \citep{Wallerstein07}, it has a rather
atypical extreme blue horizontal branch (EBHB) in addition to a blue horizontal
branch (BHB), and a very populated red horizontal branch (RHB).  \citet{Rich97}
investigated whether this could be the result of dynamical interactions in the
core enhancing the BHB via stripping of the envelope mass.  However they  found
no difference in the radial distribution of the RHB and BHB stars and took this
as evidence against the dynamical argument.  Also intriguing are the blue hook
stars (hotter and less luminous than EBHB stars), investigated in
\citet{Dalessandro08}.  \citet{Noyola06} report evidence for a modest central
cusp in the surface brightness profile and \citet{Lanzoni07} see a modest
central surface density cusp in the inner arcsec, which are interpreted by
those groups to be evidence for a central intermediate-mass black hole (IMBH)
of mass $5.7\times 10^3\msun$.  \citet{Cseh10} show that their radio
nondetection gives an upper limit to the mass of a central IMBH that is less
than half the mass predicted by \citet{Lanzoni07}.  Recently
\citet{Lutzgendorf11} used stellar kinematics to argue for a central IMBH with
a mass of 17 $\pm9\times 10^3 \msun$.  \citet{Miocchi07} suggests that tidal
stripping of red giant envelopes by this putative IMBH may be producing the
significant population of EBHB stars. 

\citet{Verbunt01} analyzed two ROSAT PSPC observations of NGC 6388 from 1991
and 1992, identifying an X-ray source at $L_X$(0.5-2.5 keV)=$6\times10^{33}$
ergs/s just outside the half-mass radius.  We do not detect a source in the
smaller of Verbunt's two error circles.  This could indicate a transient was
active during the ROSAT observations, and faint in quiescence during the
\chandra\ observations.  It also could indicate that Verbunt's ROSAT
astrometry was incorrect, since the luminosity is close to our measured total
$L_X$ for the cluster.   Less sophisticated X-ray analyses of our Chandra
observation and a 2003 XMM observation, with the goal of constraining the
presence of an intermediate-mass black hole at the cluster center, were
performed by \citet{Cseh10,Nucita08}.  

Evidence that NGC 6388 contains at least one transient X-ray binary (in
agreement with predictions of its high stellar encounter rate) was provided by
IGR J17361-4441, a hard X-ray transient located in NGC 6388. It was detected by
INTEGRAL on August 11, 2011, with follow up observations by Swift, RXTE, XMM,
ATCA, and \chandra.  It had a peak output in the 0.5-10 keV energy band of
$L_X\sim5\times10^{35}$ ergs/s \citep{Bozzo11,Nucita12}.  A short \chandra\
target of opportunity observation provided a precise position, inconsistent
with any of the cluster X-ray sources we present below, allowing an upper limit
on the transient's quiescent luminosity of $10^{31}$ ergs/s \citep{Pooley11}.
In the last Swift observation of Nov.  4, 2011 (before solar proximity forbade
observations), this transient was still active at similar luminosities, but the
X-ray emission from the cluster was consistent with the normal quiescent output
by Jan. 29, 2012 \citep{Bozzo12}.  

\section{Data \label{data}}

\subsection{\chandra\ Observations \label{chandra}}

\begin{figure}
\plotone{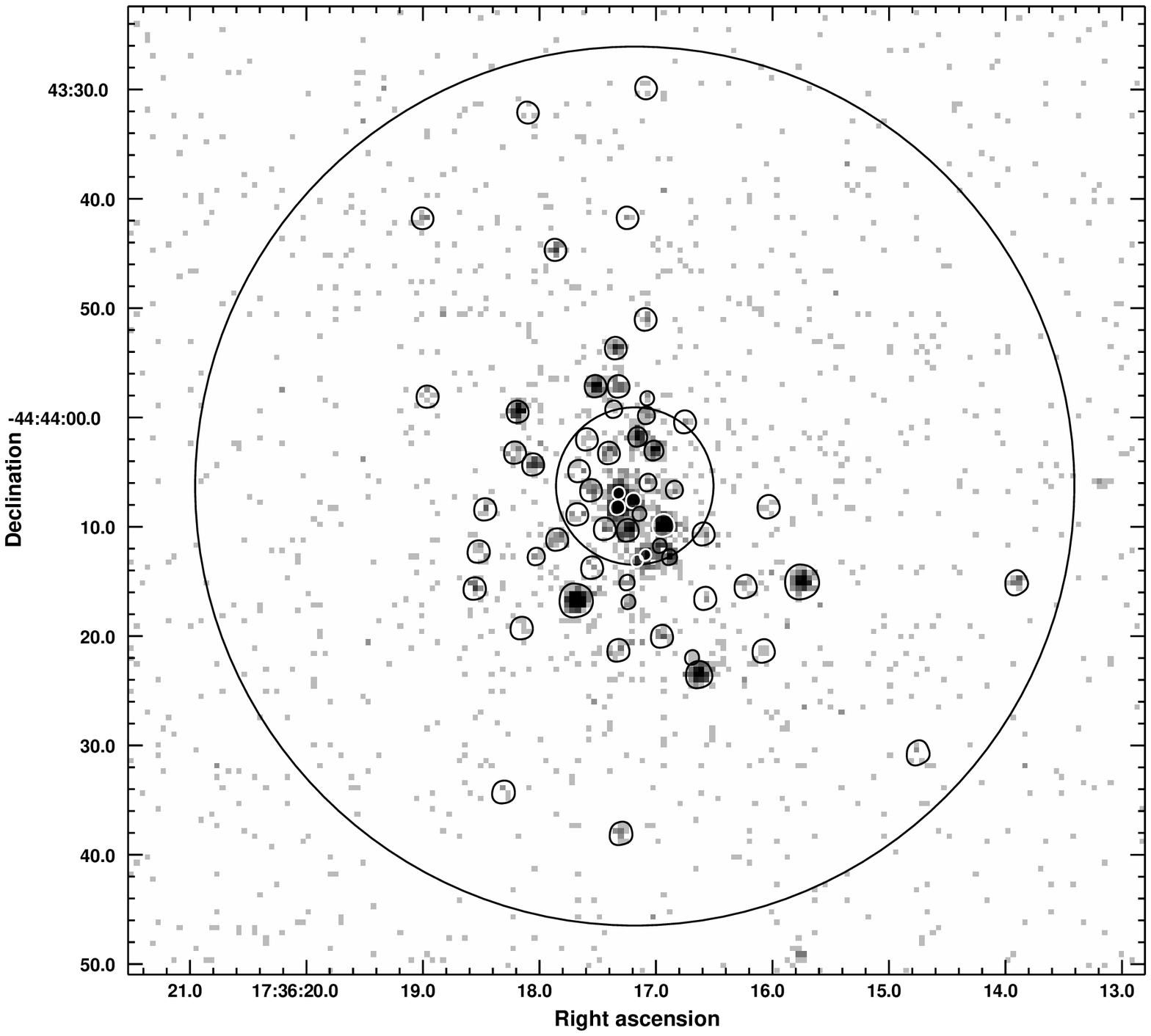}
\caption{ 
Chandra ACIS-S X-ray image of NGC 6388 (0.3-6 keV), with the core and half-mass radii indicated (7.2 arcsec and 0.67 arcminutes respectively, from \citet{Harris96}, 2010 version),
and the extraction regions for the sources within the half-mass radius of NGC 6388 marked.  
} \label{fig:halfbox}
\end{figure}

\begin{figure}[p]
\plotone{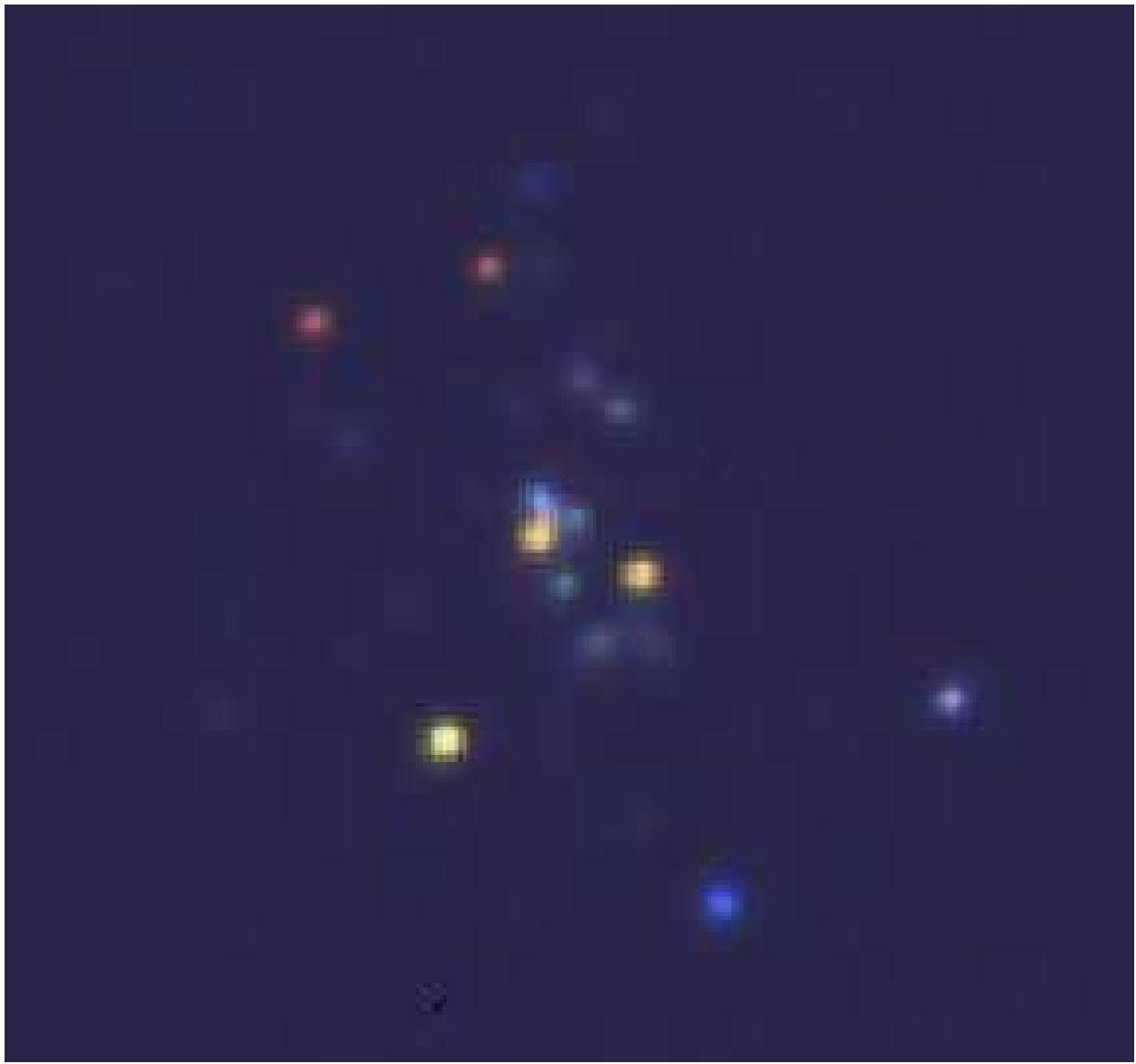}
\caption{ Representative color X-ray image of the central regions of NGC 6388 (45'' $\times$ 45''). North is up and east is to the left. Images in 0.3-1.2 keV, 1.2-2 keV, and 2-6 keV were overbinned by a factor of 2, smoothed (using CIAO tool csmooth) using a 0.5'' scale, and combined using the CIAO tool dmimg2jpg. Likely qLMXBs can be clearly identified by their yellow or red colors while the unusually blue color of CX4 stands out, at bottom right.
} \label{fig:color}
\end{figure}

We obtained a 45~ks ACIS-S\footnote{Advanced CCD Imaging Spectrometer/S-Array}
observation of NGC~6388 on 2005 April 21/22, with the center of the cluster
placed on the S3 chip. We chose the S3 chip for its high sensitivity and
spectral resolution.  We used the timed exposure mode with the very faint
telemetry format.  The moderate reddening of NGC 6388 ($E(B-V)=0.37$,
\citealt{Harris96} (2010 version)) gives a moderate hydrogen column density,
$N_H\sim2.2\times10^{21}$ cm$^{-2}$.  A detection limit of 2 counts, for the
45~ks ACIS-S exposure, corresponds to a flux of $f_X \approx 4
\times10^{-16}~\ergsc$ in the 0.5--6.0~keV energy band for a power-law spectrum
with a photon index of 2.  Assuming a distance of 9.9 kpc (\citealt{Harris96}
(2010 version)), this limiting flux corresponds to a luminosity of $L_X \approx
5\times10^{30}~\ergs$. We are complete to $\sim5$ counts, except in the core
and near bright sources. We are likely complete to $\sim10$ counts except very
near to bright sources. In Fig. \ref{fig:halfbox} we plot the location of the
identified X-ray sources on the X-ray image. We constructed a false-color
representation of the sources, Fig.  \ref{fig:color}, where the variety of soft
and hard X-ray sources is easily noted.

\subsection{\hst\ Observations \label{hst}}

\begin{figure}[p]
\plotone{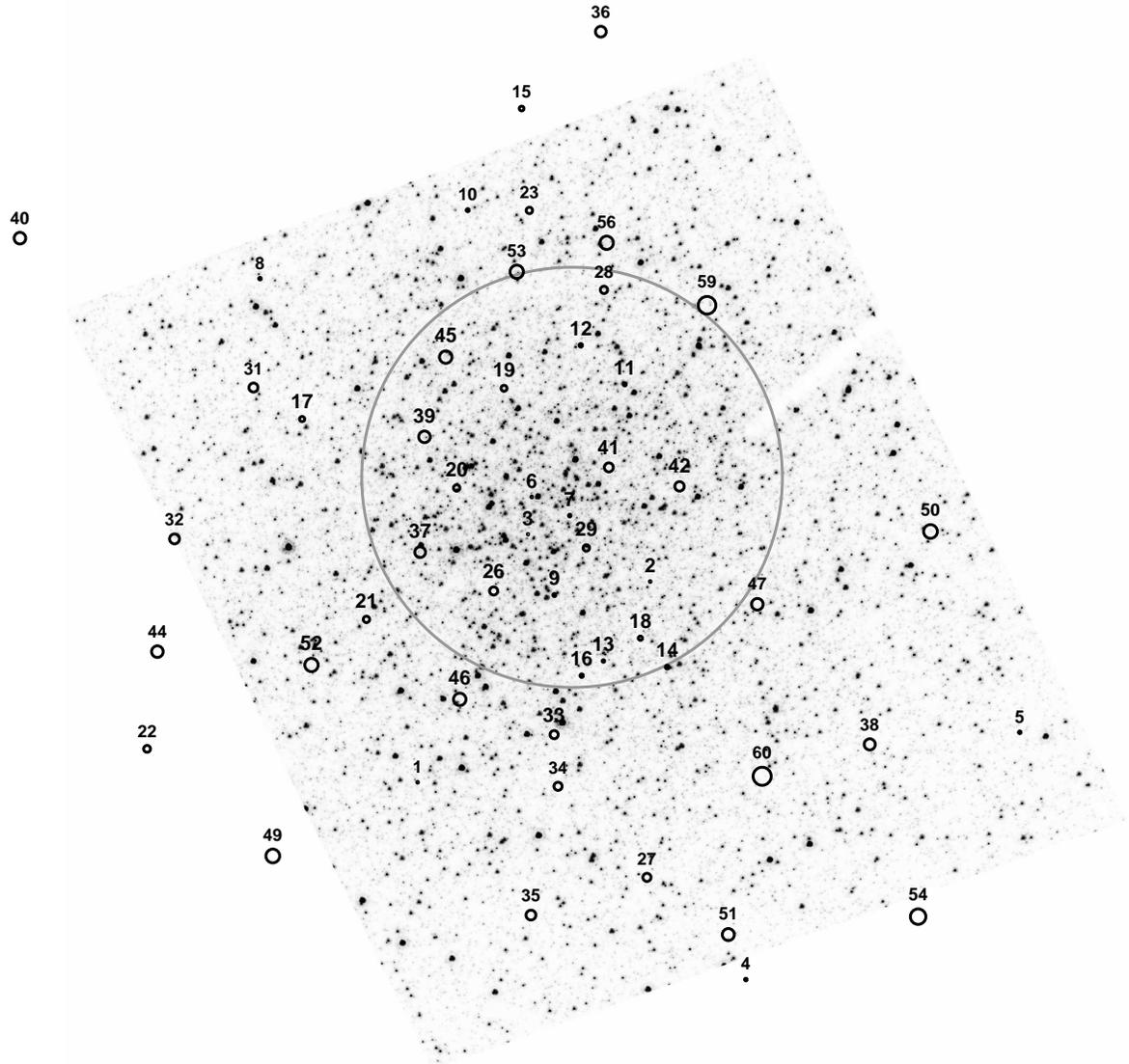}
\caption{ 
Positions of bore-sight corrected X-ray sources overlaid on the $U_{330}$ image of
NGC 6388 from 2006. North is up and east is to the left. Error circles are 1-$\sigma$ errors from determination of
X-ray source positions; no additional error is included. The large gray circle indicates the core radius of 7.2 arcsec.
}\label{f:images}
\end{figure}

Our optical analysis is based on ACS/HRC\footnote{Advanced Camera for
Surveys/High Resolution Channel} imaging obtained in 2003 and 2006 through the
F555W (hereafter $V_{555}$) and F814W (hereafter $I_{814}$) filters (GO-9835
and G0-10474, PI: Drukier) and imaging obtained in 2006 through the F330W
(hereafter $U_{330}$) filter (GO-10350, PI: Cohn). We refer to the filters used
in a way that suggests similarity to the Johnson system, but with subscripts to
indicate that the obtained data is not transformed to that system, especially
true in the F330W filter, which is both bluer and narrower than the Johnson U
filter.  Table~\ref{t:hst} lists the data sets, dates, filters, and exposure
times.  The observations used heavy dithering, with integer pixel plus subpixel
offset, in order to allow for bad pixels and to increase the effective
resolution through drizzle reconstruction of the frame.  Rather than using the
standard {\small STSDAS}~\emph{multidrizzle} algorithm, we used images created
through a similar procedure, applied by J. Anderson.  The resulting
drizzle-reconstructed $U_{330}$ image is shown in Fig.~\ref{f:images}.  We note
that while the dithering substantially improves the effective resolution by
about a factor of two, the core region of NGC~6388 remains very crowded, with a
considerable amount of psf overlap.

\section{X-ray Analysis and Results}

We reprocessed the \chandra\ data using CIAO 3.2, CALDB v.3.0.2, and standard
procedures\footnote{http://cxc.harvard.edu/ciao/threads/all.html}. This
involves identifying and excluding bad pixels, utilizing information kept in
the VFAINT telemetry format to reduce background,  using accurate
time-dependent gain and CTI corrections, removing pixel randomization, and
selecting for grade and status.  No background flares were seen, so we used the
full 45.16 ks of live observing time.

\subsection{X-ray source identification}

\begin{figure}[p]
\plotone{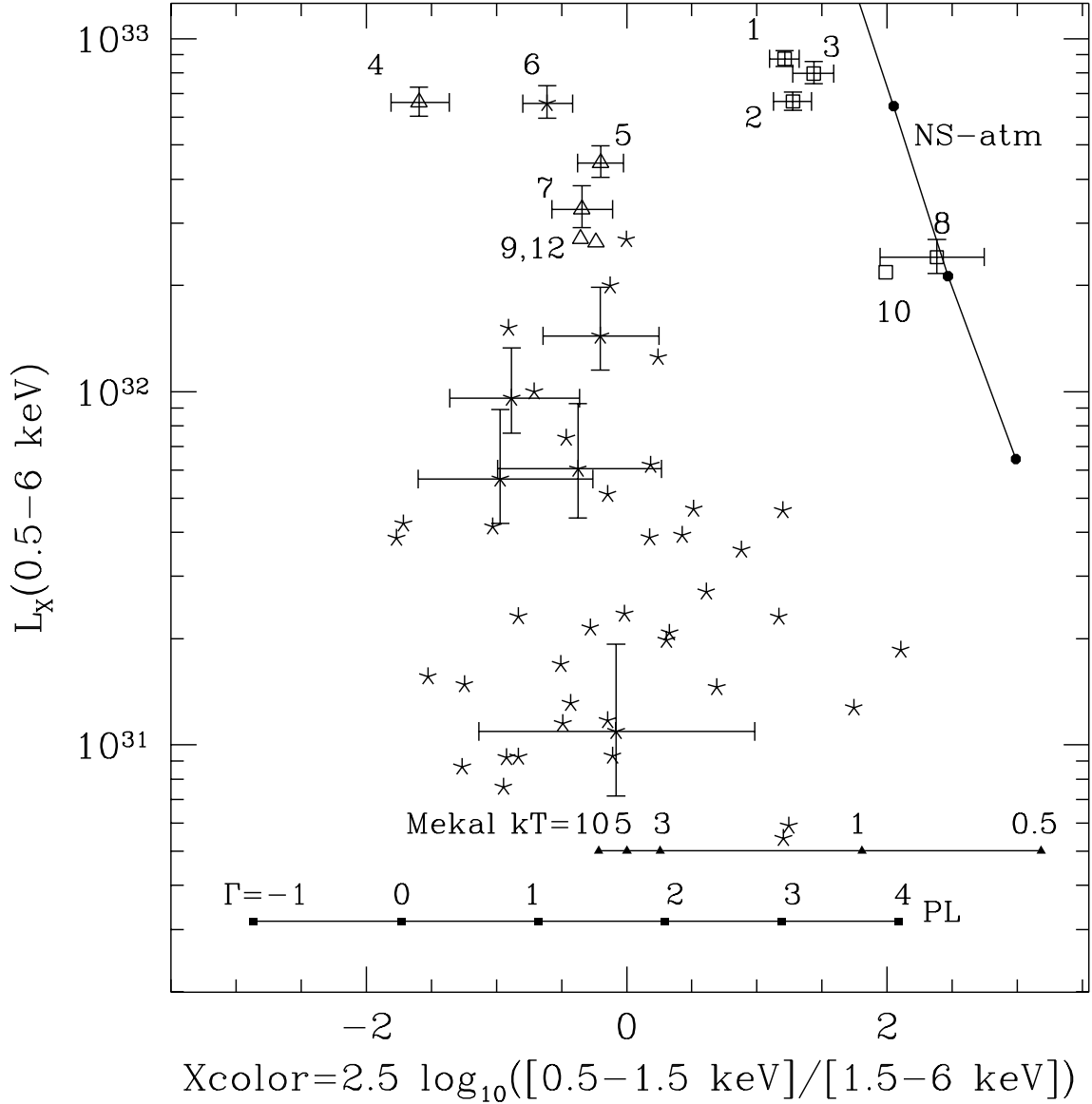}
\caption{ 
X-ray CMD of NGC 6388, plotting X-ray luminosity (0.5-6 keV, corrected for
extinction of $N_H=2.04\times10^{21}$ cm$^{-2}$) against an X-ray color for
sources within the half-mass radius of NGC 6388.  X-ray luminosities are
computed using a 6 keV MEKAL spectrum, or an average of a MEKAL spectrum and a
$10^6$ K hydrogen-atmosphere NS spectrum, to convert photon fluxes into energy
fluxes for each of 7 narrow bands, and summed.  The open squares (colored red in the online version) indicate
likely qLMXBs, and open triangles (colored blue in the online version) indicate CV candidates with optical counterparts (5,7,9, and 12) or identified from X-ray properties (4).
}\label{f:XrayCMD}
\end{figure}

We used the CIAO wavdetect program on \chandra\ images of the cluster within
the half-mass radius, in the 0.3-6 and 0.3-1.2 keV energy ranges, to detect 61
sources.  The source positions were refined, and net counts and spectra
extracted, using the IDL package acis\_extract (Broos et al. 2010); see Table
3.  Using the net counts, and a conversion from counts to flux for each of 7
narrow bands, we computed an X-ray color-luminosity diagram (Fig.
\ref{f:XrayCMD}). Five sources (marked as squares) have relatively soft X-ray
colors while being relatively luminous ($L_X>10^{32}$ \ergs).  We plot the
expected positions in this diagram of a 10-km radius, 1.4 \msun neutron star (NS) with low
B-field and a hydrogen atmosphere (NSATMOS, Heinke et al. 2006).  The five soft
sources are consistent with this line, or slightly harder (which may indicate a
spectrum containing both a NS atmosphere and a harder component). By comparison
with similar diagrams in other globular clusters \citep{Heinke03,Pooley06}, we
infer that the harder sources (consistent with e.g. a power-law of photon index
1.5) are predominantly CVs. Below we identify evidence for the CV nature of
five of them (labeled with blue triangles).

\subsection{X-ray Spectral Analysis}

\begin{figure}[p]
\plotone{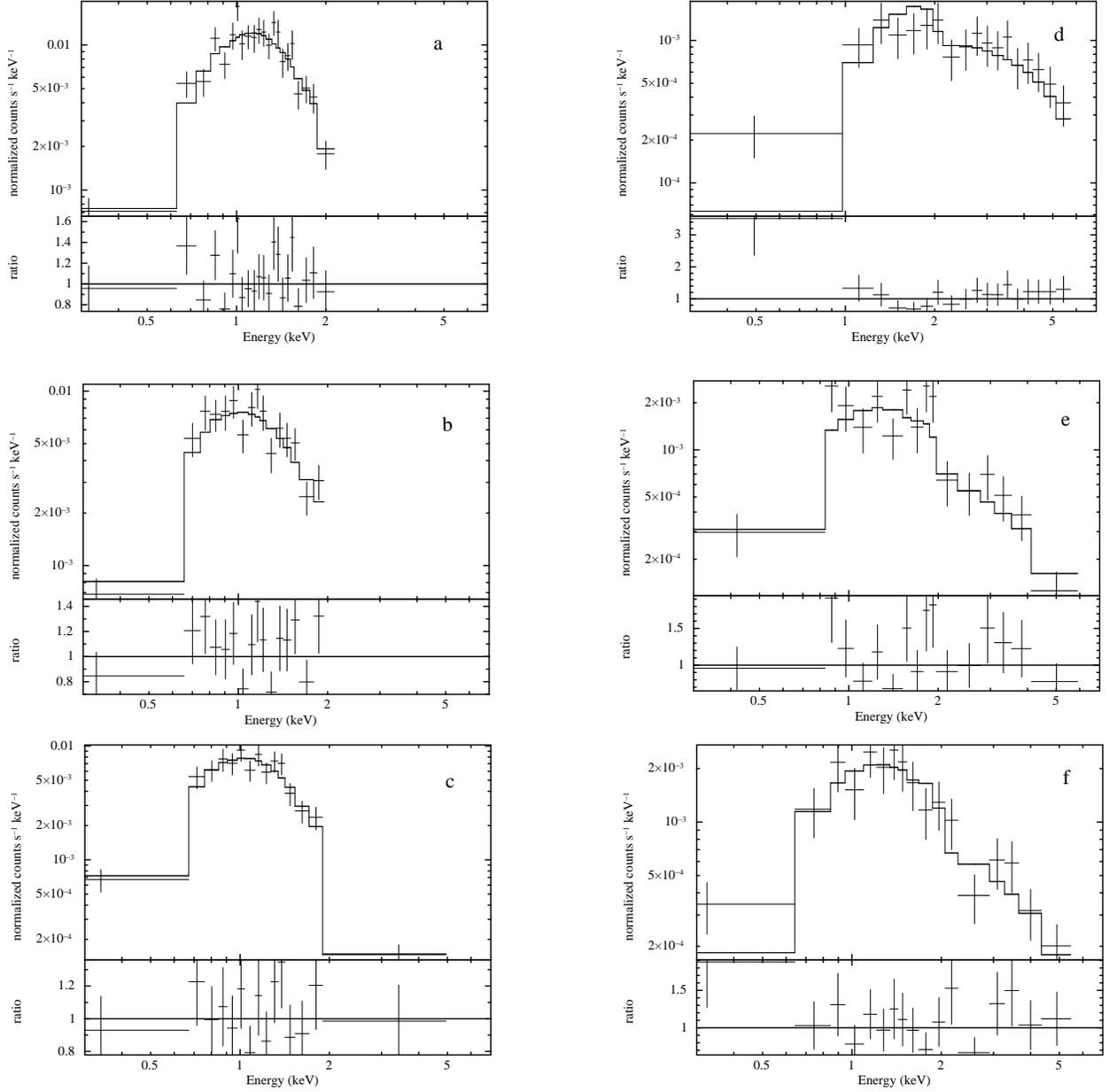}
\caption{
X-ray spectra (top, data; below, ratio of data/model) of six of the brightest
X-ray sources in NGC 6388 (a: CX1; b: CX2; c: CX3; d: CX4; e: CX6; f: CX5).  We
show here binned ($>$10 counts/bin) spectra, though our reported spectral fits
use the unbinned data and the C-statistic.  Sources CX1, CX2, and CX3, which we
argue are quiescent LMXBs, are plotted with absorbed NSATMOS+ power-law fits.
Sources CX4, CX5, and CX6 are plotted with absorbed MEKAL fits.  Note that CX4
shows an excess at low energies, arguing for a partial covering model.
}\label{f:Xrayspectra} 
\end{figure}

Below we discuss X-ray spectral analysis of those sources with more than 80
counts.  Our principal objective was to identify which sources have spectra
typical of quiescent LMXBs, vs. which sources have spectra more commonly
associated with cataclysmic variables.   We used spectral fits of a power-law,
MEKAL (hot thermal plasma, with the cluster abundance of -0.6, or 25\%;
\citealt{Liedahl}) which is often appropriate for fitting CVs, and NSATMOS +
power-law (typical of quiescent LMXBs, \citealt{Heinke06}). We performed
spectral fits in which we binned all spectra, using 20 counts/bin for brighter
sources and 10 cts/bin for sources below 150 counts, and directly on the
unbinned data using the C statistic.  The results from these fits were similar,
but the unbinned spectral fitting allowed better constraints, particularly on
the fraction of flux in the (above $\sim$2 keV) power-law component in
NS+power-law spectral fits.  Results from the unbinned spectral fits are
presented in Table \ref{t:spectra}, while example binned spectral fits are
presented in Fig. \ref{f:Xrayspectra}.  For all fits we imposed a minimum $N_H$
of $2.2\times10^{21}$ cm$^{-2}$, the value inferred from the reddening.

\subsubsection{Strong candidates for quiescent LMXBs}  

Five sources (CX1, CX2, CX3, CX8, and CX10; Table \ref{t:spectra}) require very
large photon indices ($>3$) when fit by a power-law model, or very low MEKAL
temperatures ($<1.5$ keV) for that model.  Spectral fits to NSATMOS models plus
power-law components give excellent fits without need of the power-law
component, which can be constrained to  $<$30\% of the 0.5-10 keV flux (for
assumed photon index of 1.5, e.g. \citealt{Cackett10}).  In four of them, the
power-law component can be constrained to $<13$\% of the flux.  We take these
spectral fits as strong evidence for a quiescent LMXB interpretation.  Combined
with the location of all five within or close to the cluster's core, and the
lack of bright optical counterparts (which argue against foreground coronal
sources), we conclude that these five are quiescent LMXBs.

\subsubsection{Hard X-ray sources}\label{S:CVs}

Most of the remaining brighter X-ray sources have spectra consistent with
typical CVs in other clusters--consistent with high-temperature ($\sim$5-10
keV) MEKAL models with absorption similar to the cluster value (Table
\ref{t:spectra}).   However, CX4 has a very hard spectrum (photon index
$\Gamma=0.9^{+0.4}_{-0.4}$). This suggests a CV nature, as CVs tend to have
hard intrinsic spectra, and have more intrinsic absorption than active binaries
or millisecond pulsars. This source cannot be well fit by MEKAL models using
standard absorption (see Table \ref{t:spectra}, Figure \ref{f:Xrayspectra}).
It can be well fit by a simple absorbed power law, but with a photon index of
0.30$^{+0.45}_{-0.39}$, unusually flat for an intrinsic X-ray spectrum. A
partial covering absorber model, commonly used for intermediate polar CVs,
provides a good (59\% "goodness") fit; using $N_H$ (PHABS) of
$2.2\times10^{21}$ plus a PCFABS component covering 92\% of the emission with
$3.9\times10^{22}$ cm$^{-2}$. This gives $kT=3.7^{+22}_{-1.5}$ keV, and
$L_X=2^{+1.2}_{-0.7}\times10^{33}$ \ergs.  The high luminosity and hard
spectrum suggestive of partial covering suggest that this source may be an
intermediate polar CV.

\subsection{X-ray source variability}

\begin{figure}[p]
\plotone{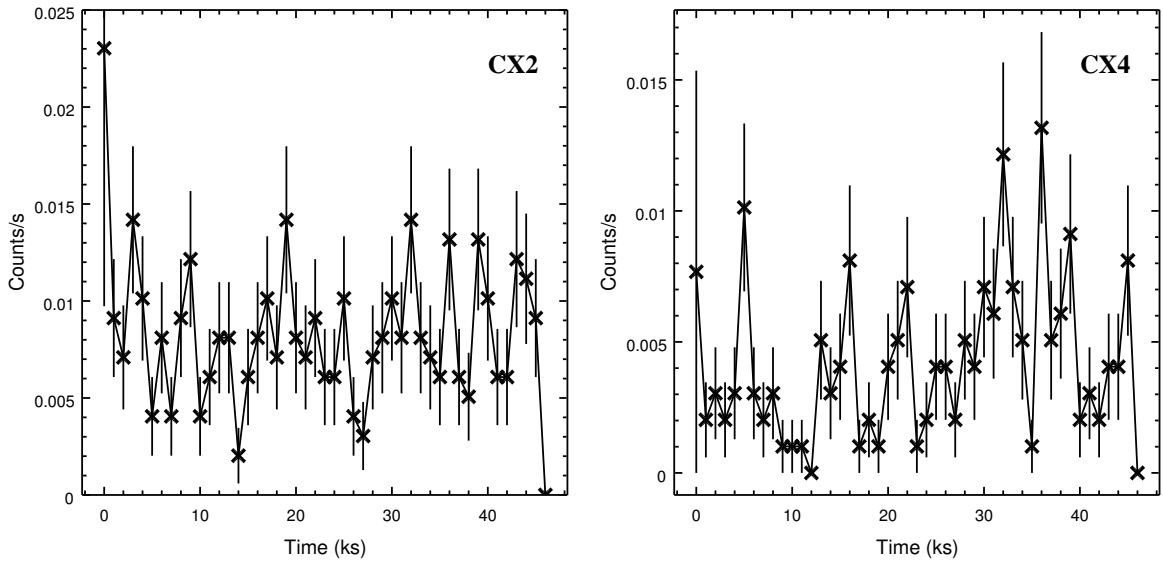}
\caption{
X-ray light curves for 2 sources, where one source (CX4) shows variation and
the other (CX2) does not. CX4 is found to vary using the glvary package in CIAO with odds ratio $10^{-1.3}$ and probability 0.043.
}\label{f:Xrayvar}
\end{figure}

We used the {\it glvary} program in CIAO \citep{Gregory92} to search for
variability in all X-ray sources with more than 30 counts.  Only one source
produced a significant detection of variability, CX4, which produced an odds
ratio of $10^{14.4}$ and probability of 1.0 of variability.  Its lightcurve
(along with that of a typical nonvariable source, CX2, with odds ratio
$10^{-1.3}$ and probability 0.043) is shown in Fig. \ref{f:Xrayvar}.  We note
that the quiescent LMXBs in NGC 6388 do not show obvious variability, as
expected if their X-ray emission arises from reradiation of heat stored in the
deep crust \citep{Brown98,Ushomirsky01}.   Some quiescent LMXBs have shown
clear variability in their thermal emission \citep{Fridriksson10, Cackett10},
but the subset of quiescent LMXBs without a strong power-law spectral component
have not yet shown evidence for such variation \citep{Heinke06}.

\subsection{X-ray Source Distribution}\label{X-raydist}

By examining the spatial distribution of the X-ray sources in NGC~6388, we may
infer the masses of the objects from which the X-ray fluxes originate (see for
instance the discussion in \citealt{Cohn10}). We compare the distribution of
the X-ray sources to the cluster parameters derived from the distribution of
stars used in finding the center of the cluster (see \S\ref{S:HSTdist}). We
expect the X-ray sources to be more centrally concentrated, and thus more
massive, assuming the populations are in thermal equilibrium.  As with previous
studies \citep{Grindlay02,Cohn10} we fit the different populations with
generalized King models:

\begin{equation}
S(r)=S_0\left[1+\left(\frac{r}{r_0}\right)^2\right]^{\alpha/2},
\end{equation}\label{E:King}

where $r_0$ is related to the core radius $r_c$ by

\begin{equation}
r_c=\left(2^{-2/\alpha}-1\right)^{1/2}r_0.
\end{equation}

Using the stars with $V_{555} < 20.0$, we find the optical distribution of NGC
6388 to be well fit by $r_c$ = 7.2 arcsec and $\alpha = -2$.  If we assume
thermal equilibrium, the slope parameter for the X-ray sources $\alpha_X$ is
related to the slope parameter for the optical distribution $\alpha_0$ by 
\begin{equation} \alpha_X = 1 + q(\alpha_0-1) \end{equation}
where $q = M_X/M_*$ is the ratio of the characteristic mass of the X-ray
sources to the mass of the objects that dominate the optical light
distribution.  We fit the King model using a maximum-likelihood algorithm. We
then applied a bootstrap method, where 1000  resampled populations are created
by resampling from the original population until the total number is reached,
replacing the chosen source back into the pool before each choice. We used the
set of these resampled populations to determine uncertainties of the best-fit
parameter values by fitting the King model to each resampled population and
then analyzing the resulting distribution of the best-fit parameter values.
The uncertainty in each parameter was set equal to half of the 68\% spread of
the best-fit parameter values about the median.  For the fits we used the
center of the cluster as determined from the optical data (see
\S\ref{S:HSTdist}).

Since the cluster lies in a region where the Galactic ridge emission is weak
(l=345.56$^\circ$ and b=-6.74$^\circ$, \citealt{Harris96} (2010 edition); see
the map in \citealt{Revnivtsev06}), the background contamination was only
determined using Eqn. 1 in \citep{Giacconi01} to estimate extragalactic
contamination.  We find that over the area covered by all the detected sources,
there are likely 2-3 background source detections.  In an effort to verify this
value, we performed an additional fit including the additional parameter $B_0$
for the background to Eqn. \ref{E:King}, as in \citet{Elsner2008}.  This gave a
result for $B_0$ that was consistent with our previously calculated value.

\begin{figure}[p]
\plotone{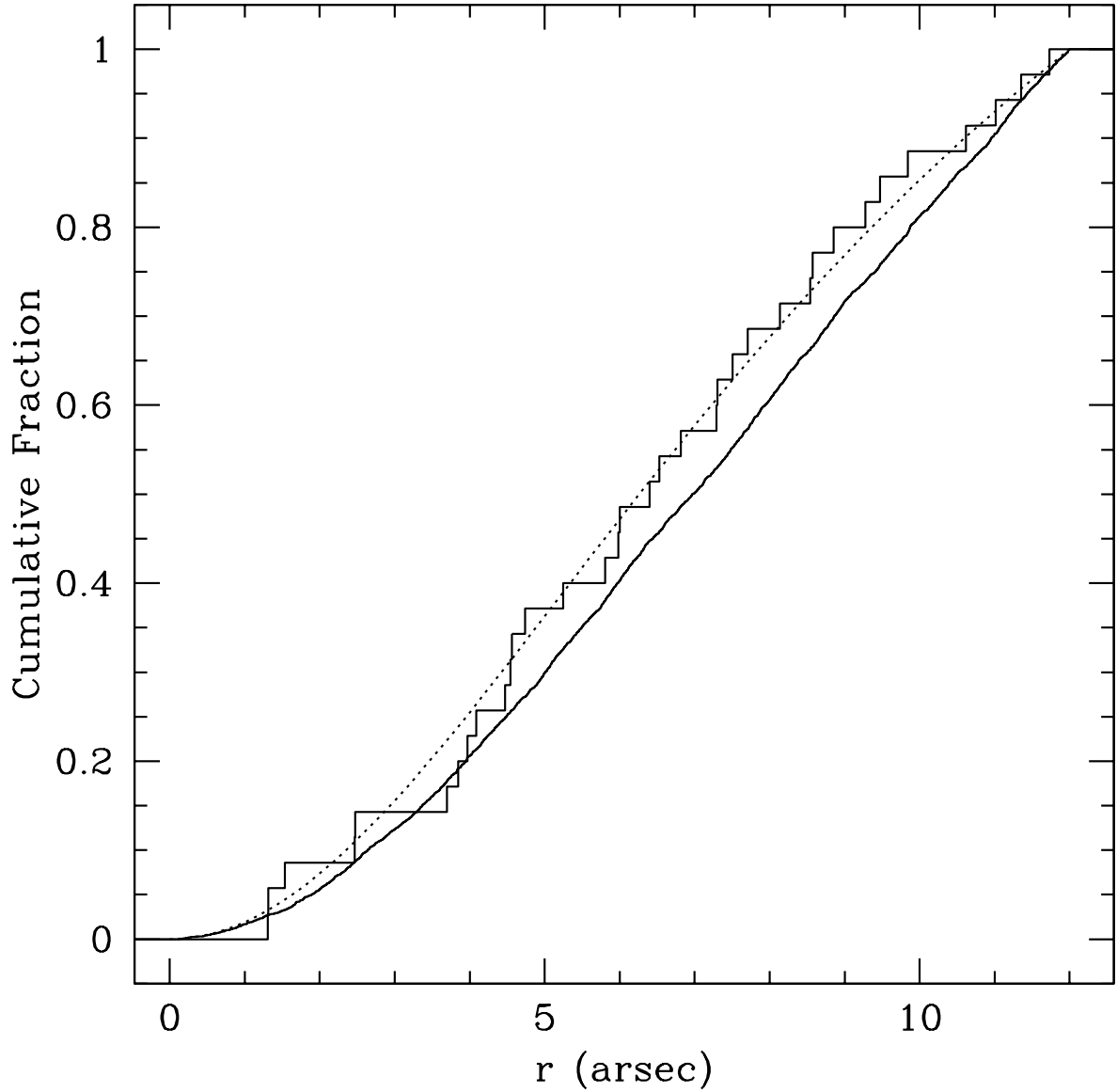}
\caption{Comparison of the radial distribution of stars brighter than V=20 (the nearly smooth line) with that of all the X-ray sources. The best-fit generalized King model for the X-ray sources is shown by the dotted curve (colored red in the online version).}\label{f:starvchanddist}
\end{figure}

The fit for the distribution of the X-ray sources by a King model profile along
with a term for background sources is seen in Fig. \ref{f:starvchanddist}. It
is plotted alongside the distribution for the stars with $V_{555}<20.0$ for
comparison. From the fits (see Table \ref{t:pops2}), we find the ratio of X-ray
source mass, $M_X$, to the typical mass of stars with $V_{555}<20.0$ to be
$q=1.51\pm0.12$.

We can infer the mass of the stars used to determine the cluster parameters by
comparison with 47 Tuc, which \citet{Catelan06} have found to be very similar
in composition and age to NGC 6388.  \citet{Heinke05} found that for the range
of stars in 47 Tuc similar to those we used to determine the optical profile
for NGC 6388, the average mass was approximately 0.88$\pm 0.05 M_\sun$. We use
the approximate value of 0.9 $M_{\odot}$, to infer a mean X-ray source mass of
$1.36\pm0.11 M_\sun$.  This is consistent with a mixture of massive cataclysmic
variables and low mass X-ray binaries. For comparison, \citet{Cohn10} reported
a mass of 1.14$\pm0.14 M_\sun$ for CVs in NGC 6397. And for 47 Tuc,
\citet{Heinke05} found a mass of 1.43$\pm0.17 M_\sun$ for all of the detected
X-ray sources, and 1.31$\pm 0.22 M_\sun$ for the identified CVs.

\begin{figure}[p]
\plotone{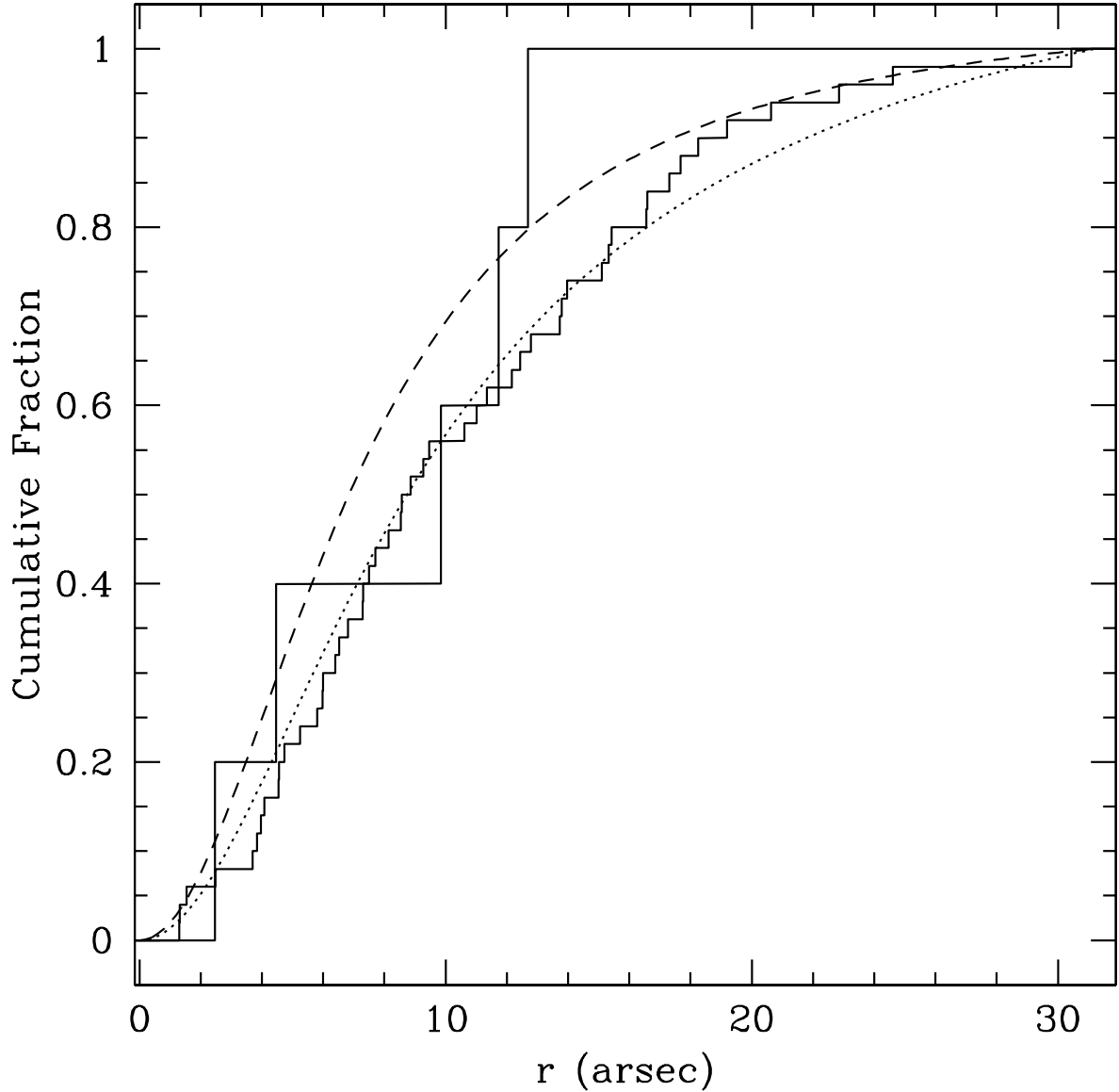}
\caption{Comparison of the radial distribution for non-qLMXBs X-ray sources (the more complete distribution curve) vs.  qLMXBs identified in this paper. For each distribution, we plot the best-fit generalized King model: dotted for the non-qLMXBs and dashed for the qLMXBs (colored red and green, respectively, in the online version).}\label{f:chanddist}
\end{figure}

\begin{figure}[p]
\plotone{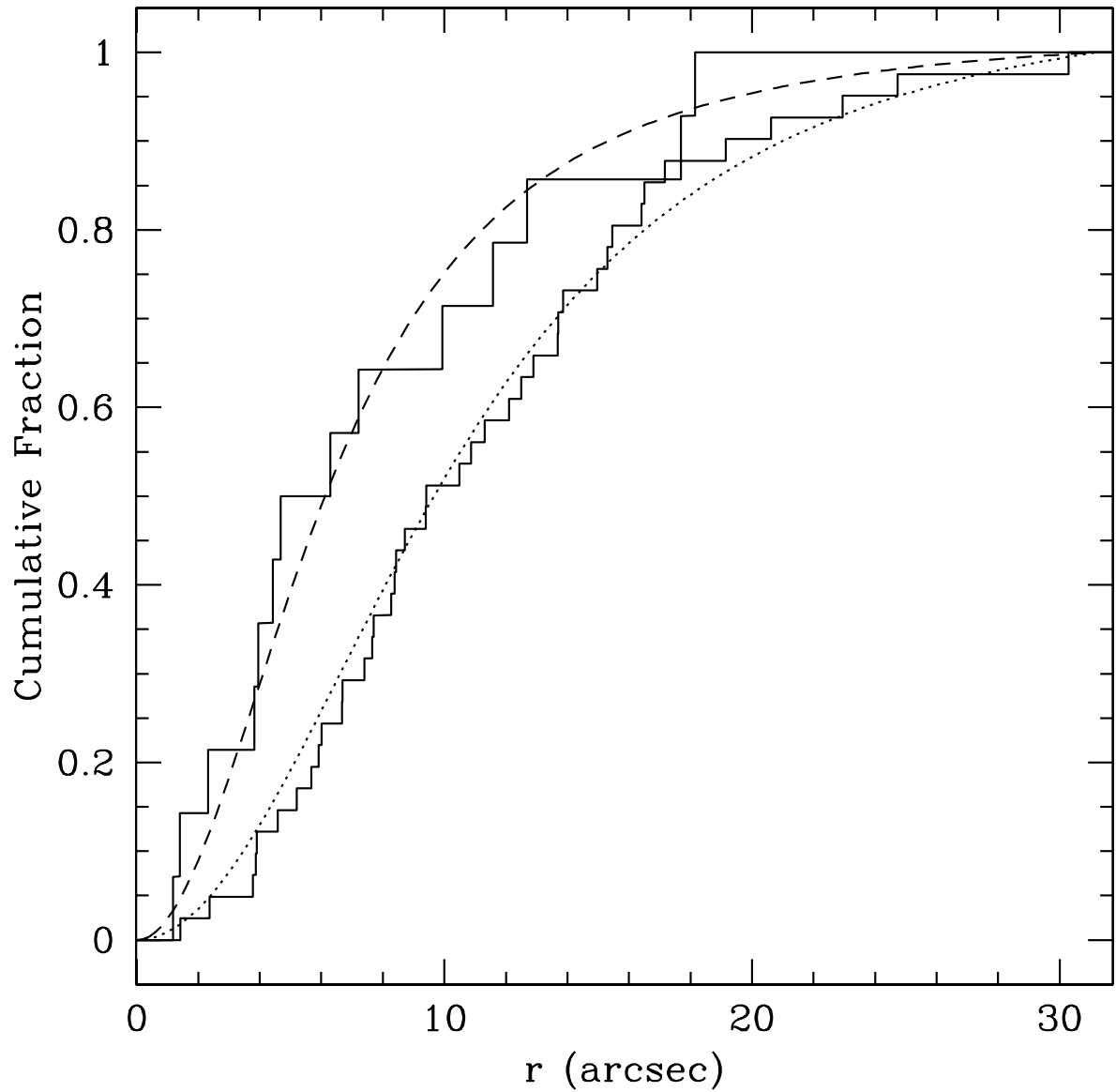}
\caption{Comparison of the radial distribution for X-rays sources with counts $>$ 40 vs. those with counts $<$ 40. For each distribution, we plot the best-fit generalized King model: dotted for the faint sources and dashed for the bright sources (colored red and green, respectively, in the online version).}\label{f:chandbrightfaintdist}
\end{figure}

We also fit the identified qLMXBs, but ignored background contamination due
to the small sample. This gave a value of $q=1.63\pm0.31$, which corresponds
to a mass of approximately $1.47 \pm0.28 M_\sun$. While the difference in mass
is not statistically significant, it is suggestive that the qLMXBs are more
massive than the other X-ray sources (see Fig. \ref{f:chanddist}). 
We also can compare the bright X-ray sources, defined as those with counts $>
40$, with the faint X-ray sources. The fits to these two populations can be found
in Table \ref{t:pops2} and their distributions are plotted in Fig.
\ref{f:chandbrightfaintdist}.  

We use a Kolmogorov-Smirnov (K-S) test on two samples to determine the
likelihood of their radial distribution being drawn from the same population of
stars.  We find they are consistent with being from the same distribution at a
probability of only 9\%. While this does not represent a high level of
statistical significance, it suggests that bright X-ray sources may be more
centrally concentrated than fainter ones, similar to that seen in NGC 6397 for
the bright CVs \citep{Cohn10}. This may indicate that dynamical interactions
scatter hard binaries into elongated orbits where they dim as they age.

\section{\hst~Data Analysis and Results}

\subsection{Astrometry}

We put the ACS/HRC frames on the ICRS astrometric grid by first obtaining an
astrometric solution for a ground-based frame, then using this frame to define
secondary astrometric standards, and finally using these secondary standards to
obtain a solution for the ACS frames.  The ground-based frame, from the ESO
archive, was taken with the Superb Seeing Imager 2 (SUSI2) on the ESO New
Technology Telescope (NTT) on 2003 May 05 with a field of view of $5.46'\times5.75'$.  The primary astrometric standards are from the second US
Naval Observatory CCD Astrograph Catalog (UCAC2), which has an RMS uncertainty
of 70 mas \citep{Zacharias04}.  Consistent with this uncertainty, the
astrometric solution for the ESO frame using 108 primary standards had an RMS
uncertainty of 72 mas.  Since the internal error of the ACS astrometric
solution using 15 secondary standards is about 13 mas, the overall accuracy of
the astrometric solution of the ACS images is 73 mas.  We took the corrected
\hst\ coordinates to be our fundamental coordinate system and applied a
boresight correction to the \chandra\ source positions as described below in
section \ref{bore-sight}.  

\subsection{The Color-Magnitude Diagram}

We used {\small DAOPHOT-ALLSTAR} to carry out PSF-fitting photometry on the two
stacked $V_{555}$ frames (2003 and 2006 epochs) and on the single stacked
$U_{330}$ frame (2006 epoch). The PSF was constructed using about 100 stars
with the least crowding in the frames, with an iterative approach to neighbor
subtraction.  We constructed the PSF by first solving for it with a total
radius that was relatively small: about twice the FWHM of the stars in the
image. After each solution for the psf, a subtracted frame was constructed to
clean the region around the chosen psf stars. A new psf with a larger radius
was determined from this cleaner frame. This process was iterated until the
change in the error to the psf fit became negligible.   A total of 28612
objects were jointly detected in the 2006 $V_{555}$ and $U_{330}$ data. 

The magnitudes were calibrated to the {\small STMAG} system using the method
described by \citet{Sirianni05}. Since the stacked images used for our PSF
analysis had an arbitrary normalization, we constructed a second set of stacked
images using \emph{multidrizzle}. We then obtained aperture photometry from a
subset of our stars from these images to the {\small STMAG} system.
Specifically, we used the psf stars we had chosen for the {\small
DAOPHOT-ALLSTAR} analysis. These magnitudes were converted to the  {\small
STMAG} system and then an offset was found between these derived magnitudes and
the previously measured psf photometry. This offset was then applied to all the
psf magnitudes in each of the filters and epochs.

\begin{figure}[p]
\plotone{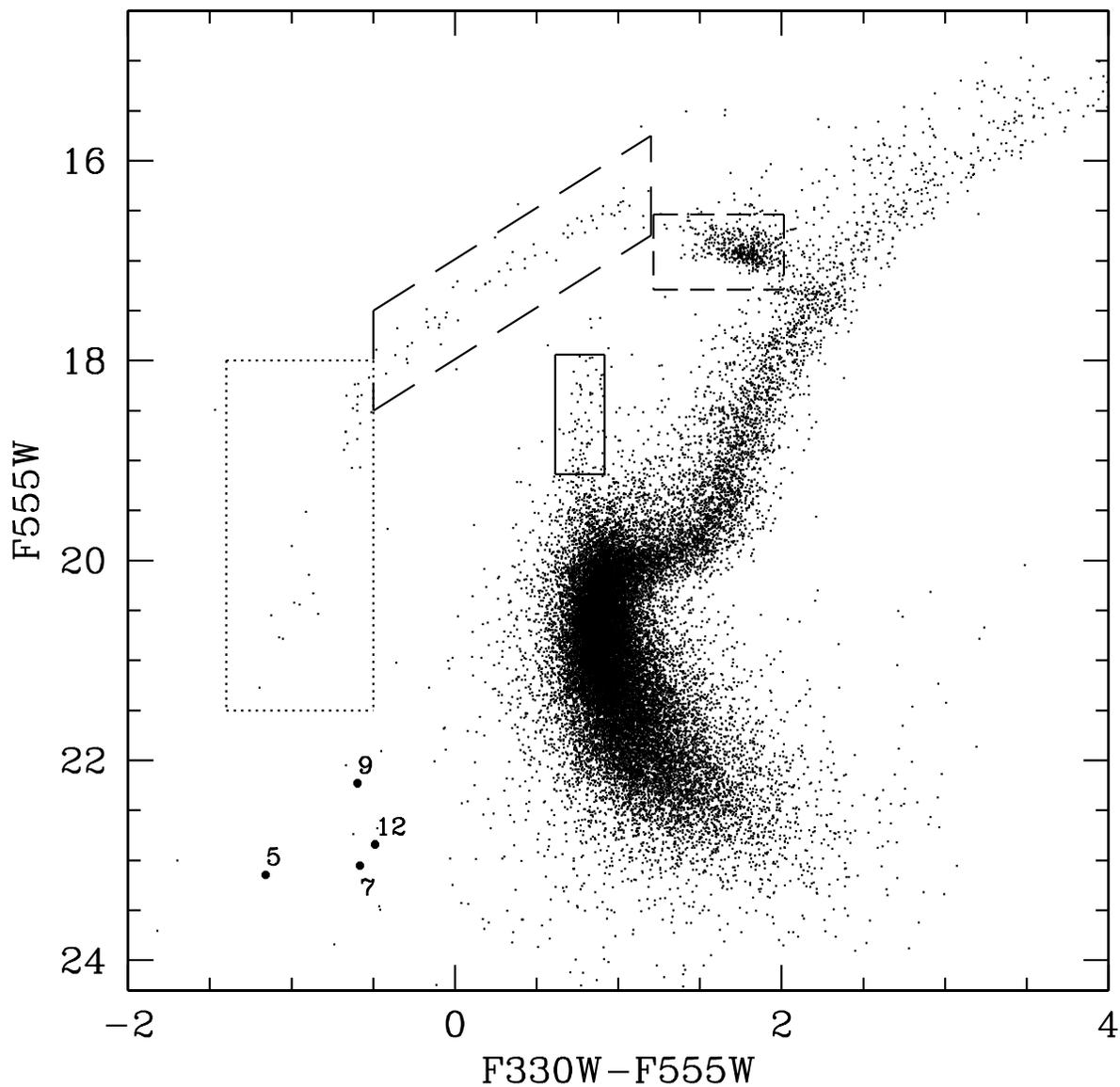}
\caption{
CMD for NGC 6388 with 4 optical counterparts labeled by their CX designations
as used in this paper. The four stellar populations we compare in table
\ref{t:pops} and fit in table \ref{t:pops2} are outlined by the limits used for those populations: RHB (short dashed, colored red in the online version); BHB (long dashed, colored blue in the online version); EBHB (dotted, colored cyan in the online version); and BS sequence (solid, colored magenta in the online version).
}\label{f:CMD} \end{figure}

Figure~\ref{f:CMD} shows the resulting ($V_{555}$,
{\mbox{$U_{330}\!-\!V_{555}$}}) CMD\@ for the 2006 epoch -- the epoch for which
both U and V band imaging are available.  The previously noted features of the
CMD, discussed in \S\ref{ngc6388}, are apparent: the sloped red horizontal
branch, the BHB, and the EBHB\@.  Also we see a large blue straggler (BS)
population, although the smearing out of the fiducial sequences makes it
difficult to select blue stragglers near the main sequence turn-off (MSTO) with
certainty.  The MSTO lies about 3.7 magnitudes below the mean magnitude of the
RHB and the MS extends about another 1.5 mag below the MSTO\@.  Of note is the
extension of the EBHB to about 1 mag below the MSTO and the presence of stars
that lie below the MSTO and which are much bluer than the MS\@.  As we discuss
in \S\ref{counterparts} below, some of these stars appear to be counterparts to
\chandra\ sources.  It is useful to compare these results to those of
\citet{Catelan06}, who have presented a deep \hst-based CMD of NGC~6388\@.
They used data from the \hst\ SNAP-9821 and GO-9835 datasets; the former was
obtained with the ACS/WFC while the latter, which was obtained with ACS/HRC for
the present program, provides higher angular resolution.  Their analysis of
these data with {\small DAOPHOT-ALLFRAME} produced a ($V$,\bi) CMD\@.  While
their overall CMD is relatively similar to ours, taking into account the
different color indices used, we note that the EBHB in our CMD extends to an
additional 2.5 magnitudes below the RHB, i.e.\ a total of 4.4 mag below the
RHB\@.  While the tip of the EBHB region is relatively sparsely populated, it
raises many questions regarding stellar evolution (see, for instance, the
discussion by \citealt{Brown10}).

\subsection{Optical Radial Distribution}\label{S:HSTdist}

The high resolution of the cluster core provides an opportunity to measure the
position of the cluster center and analyze the radial distributions of the
different populations of stars seen in the optical.  To avoid problems with
completeness, we use only stars brighter than 20 in $V_{555}$ to determine the
optical center of  the cluster in the ICRS reference frame. This sample is used
specifically to mirror the evaluation of the cluster center as described in
\citet{Lanzoni07}. The centering method involved  iterative centroiding over a
circular aperture of radius 9$''$. Using this sample of stars we find a center of
RA = 17$^h$36$^m$17.185$^s$ $\pm$ 0.12$''$ dec = -44$^{\circ}$44$'$06.44$''$ $\pm$
0.12$''$.  To find errors for the center, we use the bootstrap method as
described in \S\ref{X-raydist}. We find that due to the small field of view for
the ACS/HRC, the center cannot be determined to better than about 0.12 arcsec.
This is because the cluster's core fills much of the field of view, giving
little leverage for determining the center.

\begin{figure}[p]
\plotone{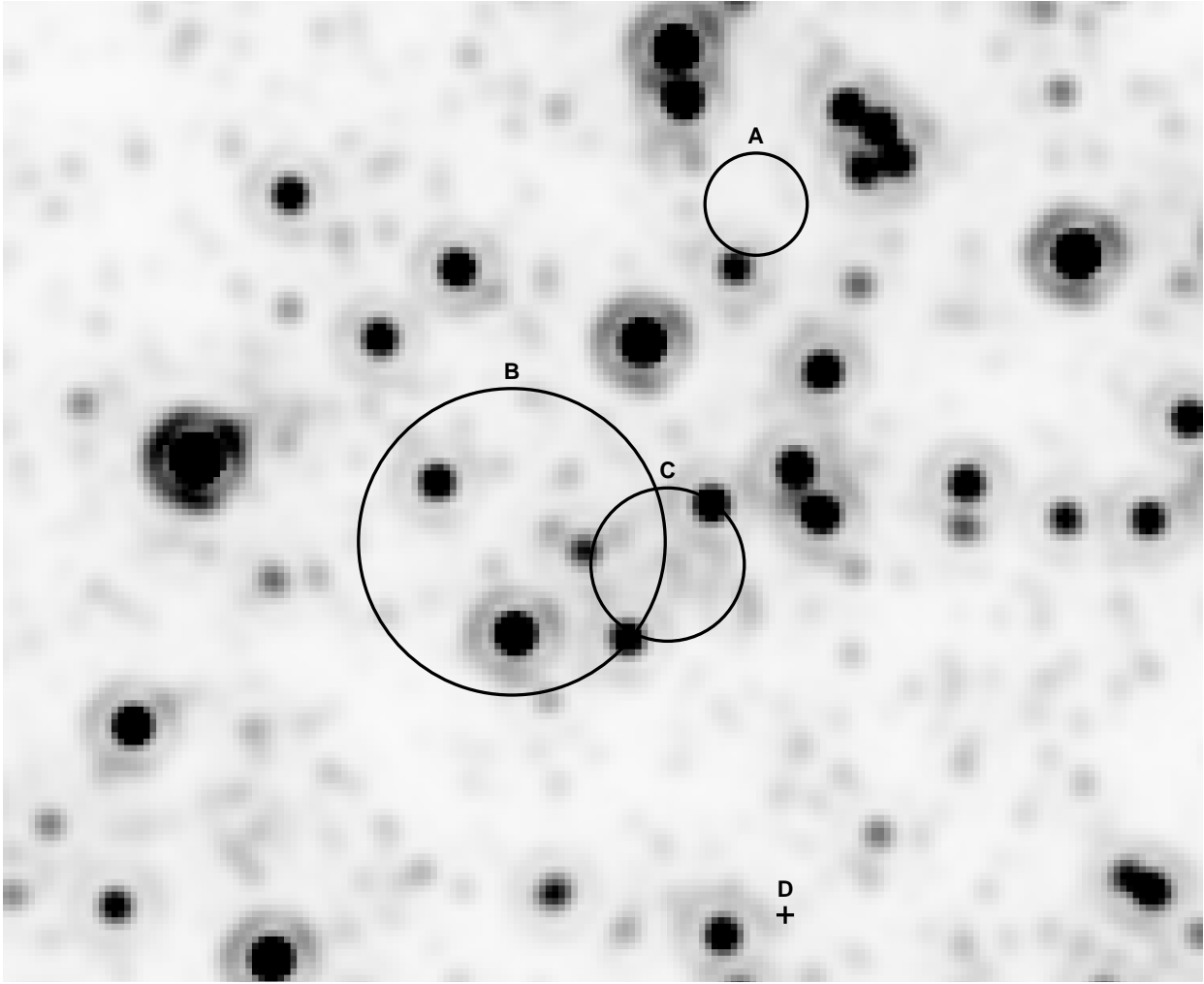}
\caption{Finding chart for the center of NGC 6388 in the $V_{555}$ filter. Error circles are 1-$\sigma$ errors on the positions as reported by the authors of A: this research, B \citet{Lanzoni07}, and C: \citet{Lutzgendorf11}. For the center labeled D from \citet{Noyola06}, no error was given by the authors.}\label{f:center_find}
\end{figure}

In Fig. \ref{f:center_find}, we compare our computed center with other recently
reported cluster centers. We find that our center differs by about
2--3 $\sigma$ from the centers determined by \citet{Lanzoni07} and
\citet{Lutzgendorf11}, lying somewhat to the north west. With our higher
resolution, we were able to count more stars than either study, which may
account for the differences. Note that the center derived by \citet{Noyola06}
was from a frame that was not corrected to a standard world coordinate system,
which (as they point out) can be significant in \hst\ fields. 

The stars used for finding the cluster center were then used to solve for the
cluster parameters using maximum-likelihood routines to fit a generalized King
model to the distribution (see Eqns. 1 and 2). Bootstrap statistics are used to
determine errors on the fit. These stars give a value of $7.1\pm0.2$ arcsec for
the core radius and $-2.04\pm0.08$ for the slope of the fall off. These are
consistent with previous solutions of 7.2 arcsec and  -2.0, respectively
\citep{Harris96}, which we use for comparison with other populations of
interest.

We fit several different stellar sub-populations in the same way as we fit the
X-ray distributions (Table \ref{t:pops2}). We see that the RHB core radius fit
is consistent with the parameters previously found for this cluster of 7.2
arcsec. We also find it is has an $\alpha=-1.92\pm0.33$, which is consistent
with the parameters for a relaxed system. A K-S test shows that the RHB
distribution is consistent with the distribution of the other stars with
$V_{555}<20.0$ (Table \ref{t:pops}). Since HB stars will typically be less
massive than the RGB stars that make up a majority of the the stars with
$V_{555}<20.0$, we assume they have not yet re-thermalized at their new masses. 

\begin{figure}[p]
\plotone{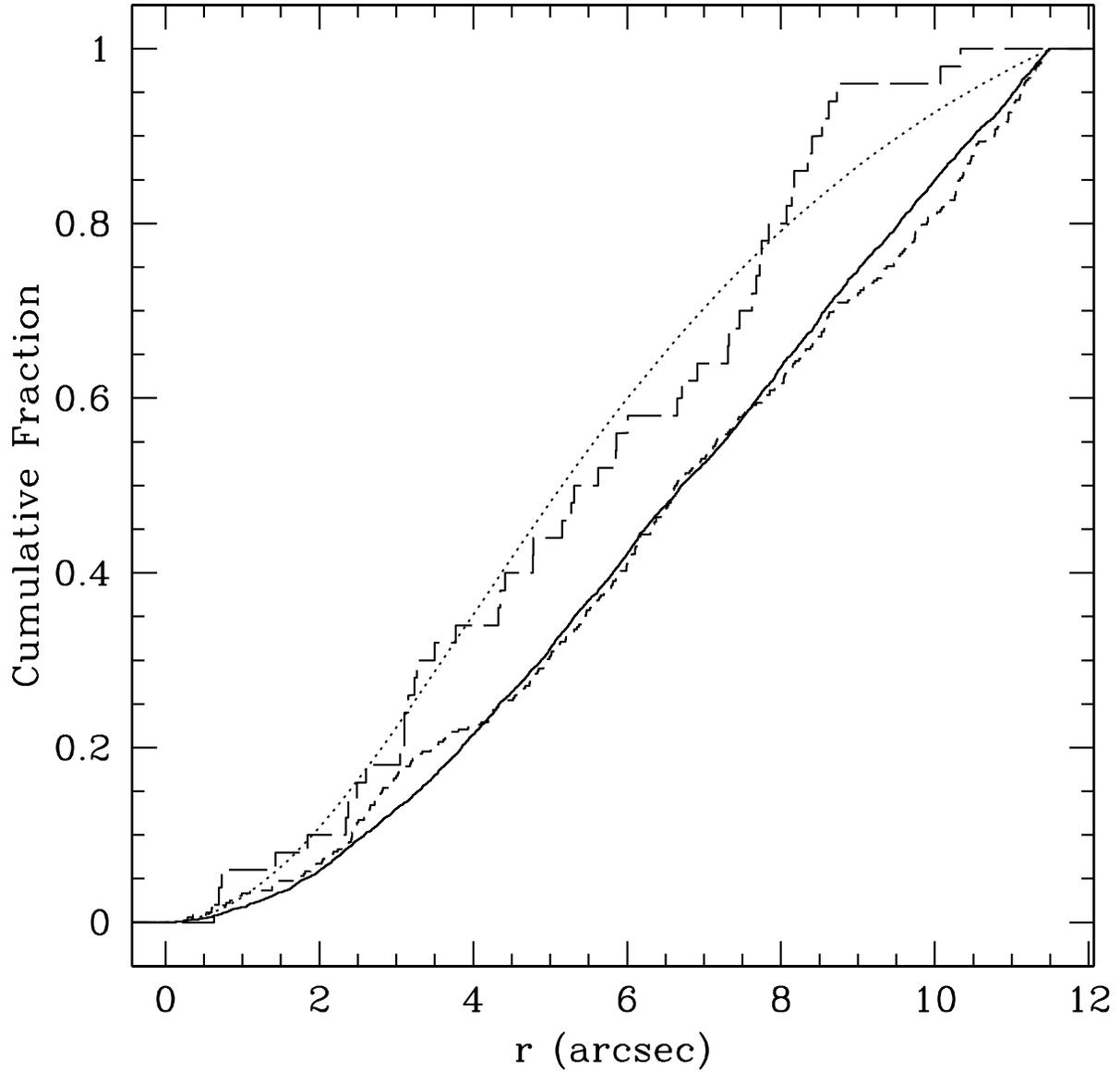}
\caption{The radial distribution of three different stellar populations. The stars
brighter than V=20 are depicted by the solid curve, the RHB stars by the short dashed
curve and the blue stragglers by the long dashed curve. For the BS, we plot the best-fit generalized King model (dotted, colored red in the online version). Note how the RHB
distribution is indistinguishable from that of the bright stars, while the BS
distribution stands out as being more centrally
concentrated.}\label{f:stardist} 
\end{figure}

Fitting the blue stragglers (BSs) in the cluster shows them to be significantly
more centrally concentrated than the stars with $V_{555}<20.0$.  We infer a
mass of 1.80$\pm0.27$ times the mass of the stars with $V_{555}<20.0$, giving a
mass of approximately $1.62 M_\sun$ for the BSs. When we plot these
distributions, we can see how the BSs stand out from the stars with
$V_{555}<20.0$, while the RHB is consistent with that population (Fig.
\ref{f:stardist}).  When we compare the BSs to the stars $V_{555}<20.0$ with a
K-S test, we find they are not the same, with a probability of less than 1\% of
being drawn from the same underlying population (see Table \ref{t:pops}).  This
is in agreement with the significant difference in the masses we find for the
two populations.

We also compare the distributions of the RHB, BHB, and EBHB stars to see if any
one is more centrally concentrated than the others as a test for the influence
of dynamical effects on the differences between these populations.  K-S tests
show that there there is no significant difference in the radial distributions
of the RHB, BHB, and EBHB. Thus we find no difference in the masses of these
three groups of stars. This result is in agreement with previous findings for
these populations \citep{Rich97, Dalessandro08}.

\subsection{Boresight Correction} \label{bore-sight}

From the CMD, a very blue star was easily picked out. This star coincides with
one of the brighter X-ray sources (source CX5), and provided a preliminary
boresight correction to the X-ray positions. After applying this correction, we
attempted to discover additional counterparts from their position in the
color-magnitude diagram and proximity to X-ray source positions. But of the
stars for which $U_{330}$ and $V_{555}$ magnitudes were obtained through the
standard procedures, there were none that could be identified in this way.
Fortunately, three additional optical counterparts were identified by blinking
the images in two bands ($U_{330}$ from 2006 against $V_{555}$ from 2003 and
2006) and searching for blue stars near the X-ray positions. Using these four
counterparts, a final correction of 0.105 arcsec in RA and -0.031 arcsec in Dec
was applied to the \chandra\ positions, well short of the 90\% pointing error
circle of $\sim0.6$ arcsec.  Once this correction was applied, we searched for
additional counterparts. In the region near source CX11, a blue object was
identified visually. Unfortunately it lies in the Airy ring of a much brighter
star, preventing accurate photometry.  We also extended our search to objects
with a red excess in the $U_{330}$, $V_{555}$, and $I_{840}$ bands in search of
millisecond pulsars and active binaries.  These additional searches yielded no
other detectable optical counterparts.

The small number of detected counterparts is not surprising considering that
the factors that predict a large number of X-ray sources in NGC 6388 also make
doing deep photometry challenging. Since the likely counterparts to X-ray
sources are most likely to be centrally concentrated, the optical images were
centered on the cluster. Because of the small field of view of the HRC on the
ACS, the moderately large core radius represents a significant fraction of the
entire field. As the core is also very dense, most of the optical counterparts
must be found against a bright background of the core stars, making sky
subtraction unreliable for objects more than about 3 mag below the MSTO. This
also results in X-ray sources somewhat outside the core (such as CX4) falling
outside of the field of view in the optical images.

\subsection{Analysis of Optical Counterparts}\label{counterparts}

\begin{figure}[p]
\plotone{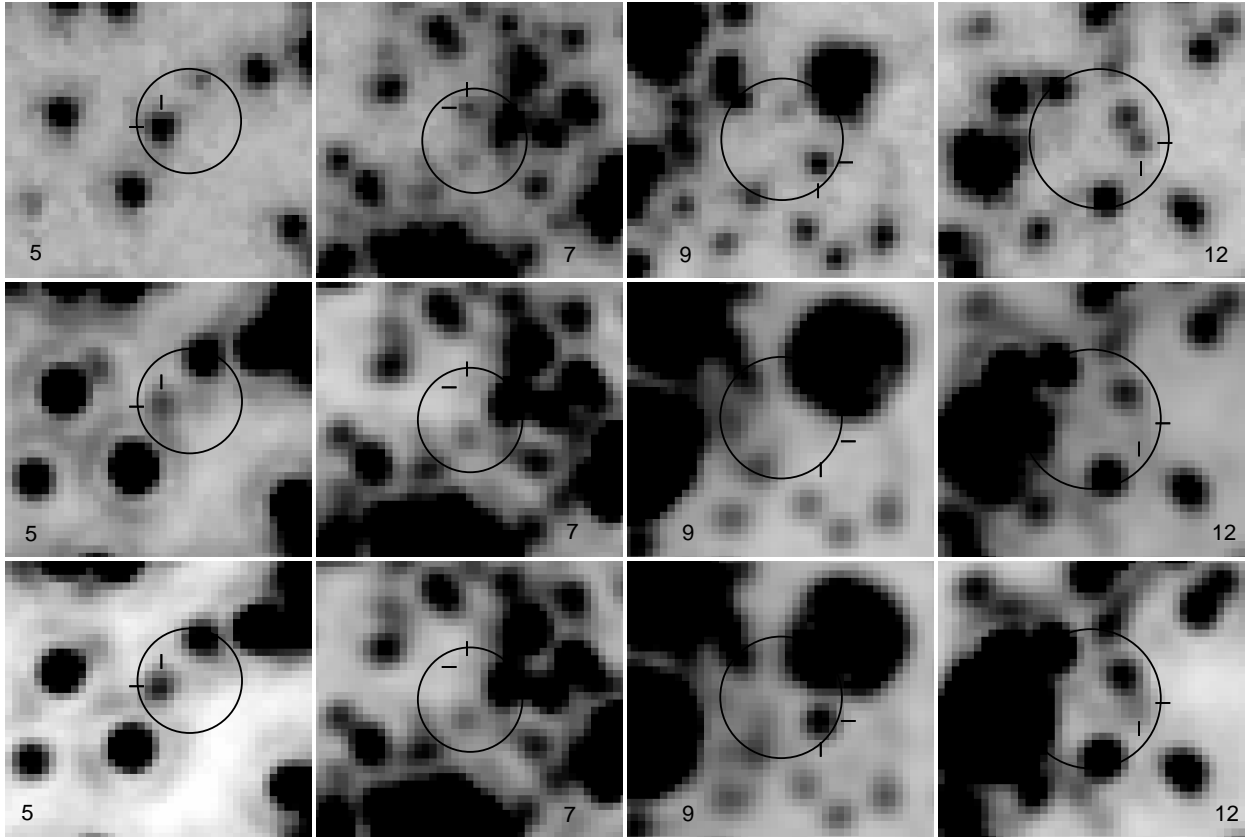}
\caption{The finding charts for the four identified optical counterparts. We show the F330W from 2006 on the top row, F555W image from 2006 on the middle row, and F555W image from 2003 on the bottom row. The circles represent two times the 68\% confidence regions for the positions of the X-ray sources. The counterparts are noted by the tick marks.}\label{f:counter_find} \end{figure}

We performed photometry for each of the identified optical counterparts using
{\small DAOPHOT-ALLSTAR} on combined, drizzled frames for each of the epochs
(Table \ref{t:counter_phot}). One can see from the CMD (Fig. \ref{f:CMD})  that
these four stars lie well to the blue of the main sequence (MS) and near the
limit of detection in the $V_{555}$ frames. For three of these stars, a light
curve was constructed using aperture photometry performed on the original,
undrizzled frames. The points were then further binned to provide a statistical
data set with which we could test the variability of the optical sources. For
the dimmest counterpart in the densest part of the field (source CX7),
measuring in undrizzled frames was not possible in either the 2003 or 2006
epoch. In this case, we drizzled and combined the data into three images. For
reference, finding charts are shown in Fig. \ref{f:counter_find}.

\subsubsection{Source CX5}

\begin{figure}[p]
\plotone{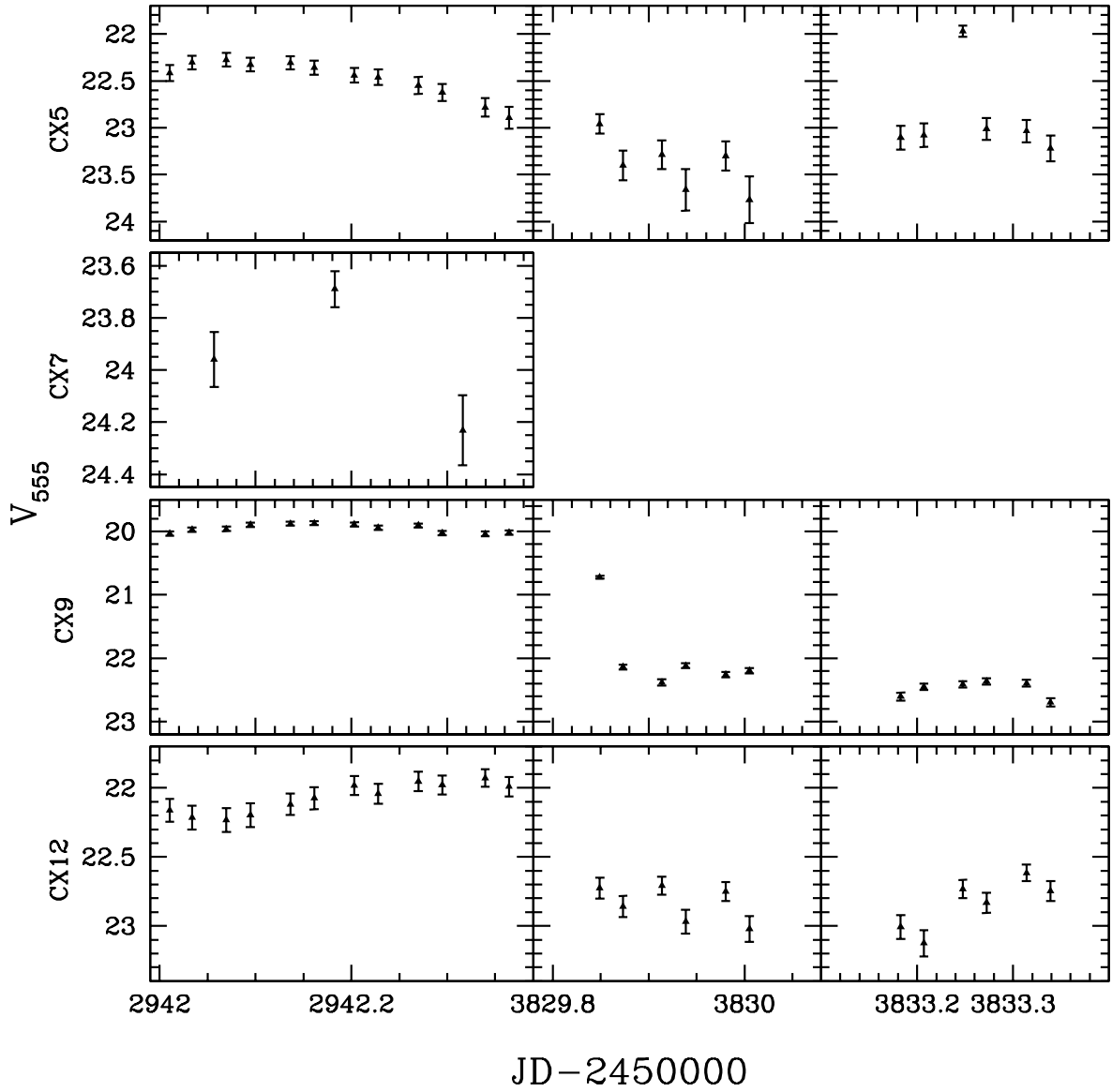}
\caption{
Light curves with 1-$\sigma$ photometric error for the counterparts CX5, CX7, CX9, and CX12 for 2003 and 2006, except in the case of CX7, which was too faint in 2006 for a light curve.
}\label{f:starLC}
\end{figure}

CX5's counterpart is extremely blue ($U_{330}-V_{555}$=-1.2). The Airy disk
from a brighter nearby star makes $I_{814}$ measurements impossible. Our
$V_{555}$ light curves for 2003 and 2006 (Fig. \ref{f:starLC}) show variability
of 0.6 and 0.8 mags (respectively), on timescales of hours. The 2003
variability suggests sinusoidal variations with P$>8$h, but the amplitude is
extremely high for ellipsoidal variability (typically 0.1 mags, e.g.
\citealt{Edmonds03b}).  The companion's average brightness drops by 0.7 mags
from 2003 to 2006.  The size of these changes may suggest dwarf nova eruptions,
which are typically 2-5 mags \citep[e.g.][]{Shara96}, although disk flickering
cannot be excluded.  Using the distance of 9.9 kpc from
\citet[][2010 version]{Harris96} and correcting for extinction, the source has a bright
absolute magnitude of 6.99.  The absolute magnitude M$_{V_{555}}$ and L$_X$ of
CX5 place it among the more luminous X-ray CVs known in clusters.
  
\subsubsection{Source CX7}

We discovered the possible counterpart for source CX7 (as well as the
counterparts for sources CX9 and CX12) by blinking the drizzled U and V images
and looking for obvious blue sources near the error circles of the X-ray
sources.  CX7's optical counterpart is clearly blue ($U_{330}-V_{555}$=-0.58),
although accurate photometry is difficult due to its being located in one of
the densest parts of the cluster. In order to construct a light curve for this
nearly blended source, we drizzled the image sets from 2003 into three images,
which clearly require hours-timescale variability (change of 0.5%$\pm$0.16
mags) (The star in the 2006 epoch was too low in signal in all the images for
this procedure to work.) Plotting these data and their associated errors shows
the star's luminosity is not consistent with a constant source (see Fig.
\ref{f:starLC}). The absolute magnitude of the counterpart in 2006 is
$M_{V_{555}}=$6.89. In view of this object's optical variability and location
in the CMD, and the X-ray source properties (\S\ref{S:CVs}),  we conclude the
optical object is a bright CV and the optical counterpart to source CX7.

\subsubsection{Source CX9}

The optical counterpart for this source is very blue ($U_{330} - V_{555}$ =
-0.60). It appears very bright (M$_{V_{555}}$=3.79) in the 2003 V data,
but is much dimmer (M$_{V_{555}}$=6.065) in 2006. It shows
variability in the $V_{555}$ of 0.2%$\pm$ 0.06
mag over the observation in 2003, and 0.6 mag in 2006 (Fig. \ref{f:starLC}).
From the curve for 2003, we find a lower limit for the period of variation of
about 8h.  Of our counterparts, this is the most convincing case for
ellipsoidal variation with its small magnitude change, indicating a possible
20-h period.  Dramatically, between the 2003 epoch and the 2006 epoch, the
optical object shows a difference in magnitude of  more than 2 mags.  This is
consistent with dwarf nova eruptions observed in globular clusters
\citep[e.g.][]{Shara96, Shara05}.  As this object also has similar optical
properties to the counterpart for source CX5, we use the same argument to
identify it as a CV.

\subsubsection{Source CX12}

This counterpart to CX12 is very blue ($U_{330}-V_{555}$=-0.49). The
counterpart is very bright with M$_{V_{555}}$=6.68 in 2006. Over the 2003
observations it shows variations of $0.2 %\pm0.08 
$ mag and in the 2006 epoch
it varies by $\sim 0.5%\pm.08 
$ mag (Fig. \ref{f:starLC}). In light of CX12's
CV-like X-ray properties, the nature of its optical variation, and its position
in the CMD, we conclude that it is consistent with a bright CV.

\section{Discussion}\label{discussion}

We find a rich population of X-ray sources in NGC 6388.  Our analysis of the
density profile for the X-ray sources shows them to be more centrally
concentrated than the brightest stars in the cluster. The inferred
characteristic mass of 1.36 $\msun$ for the X-ray sources suggests that many of
them are likely to be compact binary stars with degenerate remnant primaries.
Many of these were likely produced in close interactions in the
cluster's large, dense core.

We examine some of the characteristics of the different populations of the
optical sources. The extensive population of blue stragglers was shown to be
more massive than the horizontal branch stars. The distinctive extreme blue
horizontal branch stars were shown to be no different in distribution than the
other horizontal branch stars, confirming the result that dynamical
interactions are not a likely cause of the large color distribution in the
cluster's horizontal branch. 

We identify five sources as bright CVs based on their X-ray luminosities and
colors, together with the detection of optical counterparts for four of them,
and the detection of X-ray variability for the remaining one.  These
counterparts are consistent with bright CVs in optical variation, color, and
luminosity.  In fact, we find at least one of the sources (CX9) to exhibit
variation on par with that exhibited by dwarf nova outbursts. That would make
it only the 15th such object found in globular clusters to date \citep[see
][]{Servillat11}. The only CV candidate without an identified optical
counterpart, CX4, was identified as a likely CV based on its X-ray properties;
an unusually hard spectrum, well-fit by partial covering absorption models,
luminosity, and variability. We find five sources that have X-ray colors,
luminosities, and spectra consistent with what we expect for qLMXBS.  These
sources are not found to vary, which is consistent with expectation for sources
without a strong power-law component.

The CVs found here are some of the brightest (in X-rays and optical light) yet
known in globular clusters, suggesting relatively high mass transfer rates.
These are likely the tip of a much larger CV population.  The limit of our
detection of CVs was $V_{555}<23.1$, for which we detected 4 bright CVs as
optical counterparts to X-ray sources. We can extrapolate the possible size of
the CV population in NGC 6388 by comparison with other clusters with bright
CVs and a detected CV population. 47 Tuc was found to have at least 22 CVs by
\citet{Edmonds03} looking to 8 magnitudes fainter than the MSTO. After
adjusting for distance and reddening, there are 4 CVs in 47 Tuc with 
magnitudes similar to those in NGC 6388. We can infer that the population of CVs in
NGC 6388 is likely as large as that of 47 Tuc. This is not surprising given
the similarities between the two clusters and that they both have some of the
highest encounter rates of any cluster (see below).

The large number of sources detected in our data is consistent with predictions
based on the encounter rate. We can estimate the encounter rate using the
approximation given in \citet{Verbunt03}, $\Gamma \propto \rho_0^2 r_c^3/v_0$.
We use values from \citet{Harris96} (2010 edition) for $\rho_0$, $r_c$, and
$v_0$ for this approximation. In Table \ref{t:collision}, we give our
calculation for this value for the clusters studied in \citet{Pooley03} and for
NGC 6388 and compare these to the number of detected X-ray sources with
$L_X>4\times10^{30}$ \ergs~in the 0.5-6 keV range.   We have normalized our
values of $\Gamma$ to the mean of the values given for the non-core collapsed
clusters in \citet{Pooley03}. We find them to be roughly consistent, with major
differences likely due to our use of the updated Harris catalog. NGC 6388's
approximate $\Gamma$ is seen to be significantly larger than that of all but
one of the other clusters. In the table we also compare the number of detected
sources reported by \citet{Pooley03} with the number detected in more recent
studies (as noted in the table comments).  The larger numbers detected in more
recent studies may be attributed to improved reduction techniques that allow
for a better separation of sources in crowded regions. Ultimately, the only two
clusters to show a significant difference are $\omega$ Cen and 47 Tuc, which are most affected by the changes in their values for $\Gamma$, a result of using updated values for the cluster parameters.

\begin{figure}[p]
\plotone{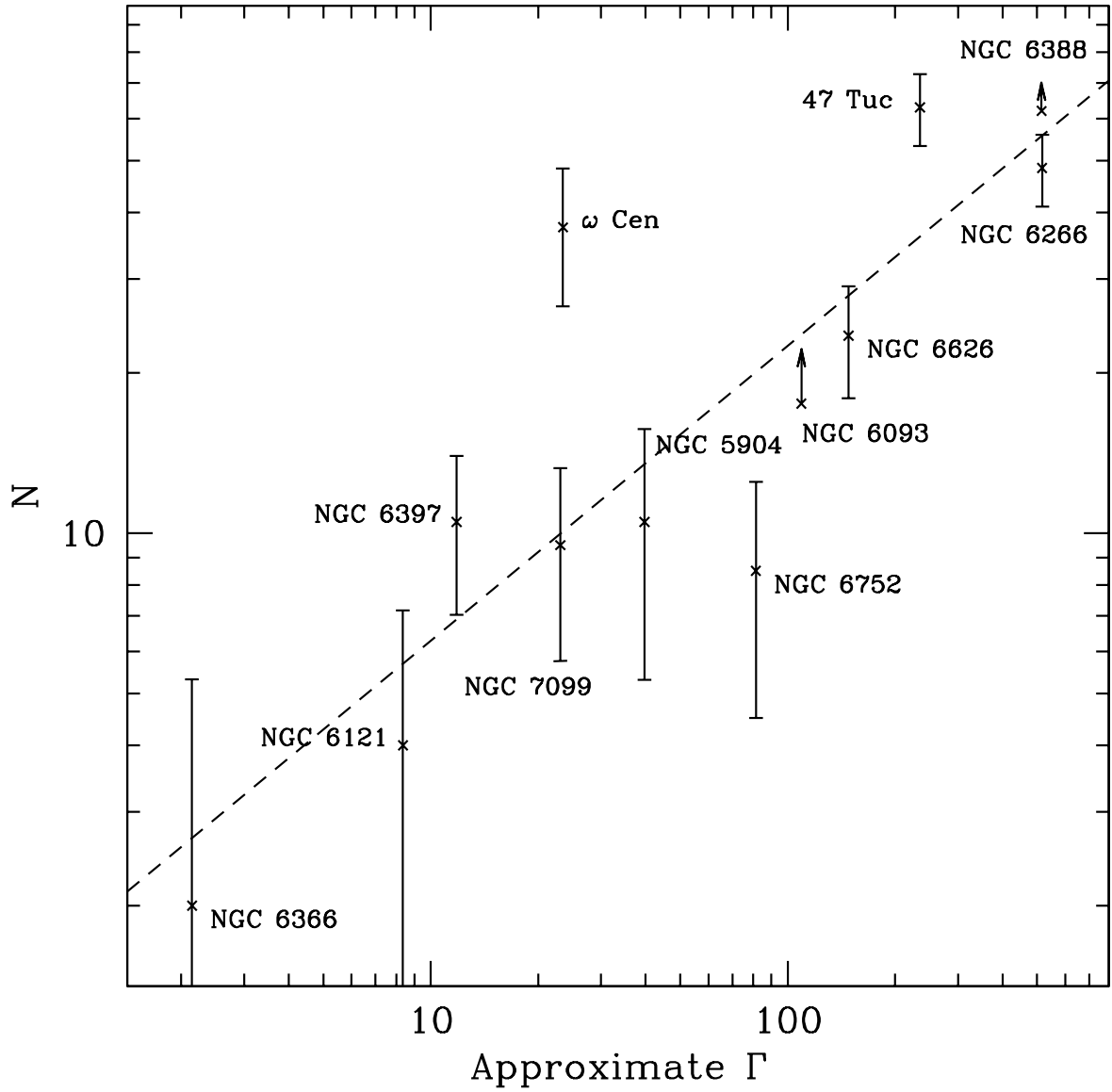}
\caption{Plot of detected number of sources vs. the approximate values of $\Gamma$ as given in Table \ref{t:collision}. The source numbers are corrected for background contamination.  The dashed line represents a least-squares power-law fit. }\label{f:gamma}
\end{figure}

In Fig. \ref{f:gamma}, we plot the number of detected sources, using the newest
values when possible and correcting for the predicted background contamination,
versus $\Gamma$. We note that detections for NGC 6388 and NGC 6093 (M80) are
not complete to L$_X>4\times10^{30}$ \ergs~ (see \S\ref{chandra}), so their
detected source numbers are lower limits (indicated with arrows).  
% REVISIT NEXT SENTENCES
We fit the relationship of $\Gamma$ and N with a power law and find it can be
described as $N\propto\Gamma^{0.55\pm 0.09}$. This corresponds with the result
of \citet{Pooley03}, where a similar trend was found with their calculated
values of $\Gamma$.  In light of this trend, the detection of X-ray sources in
NGC 6388 is a confirmation of the prediction that its large, dense core
produces a large number of close stellar interactions, which in turn
accelerate the production of close binary systems.

\acknowledgements
This research is supported by NASA grants GO5-6045X and HST-GO-10350.01-A to
Indiana University.  COH and SAB thank NSERC, and COH thanks the Alberta
Ingenuity New Faculty program, for funding.

\clearpage 

\bibliography{n6388_rev1}

\clearpage

\newcommand{\nd}{\nodata}\begin{deluxetable}{lllr}
\tabletypesize{\normalsize}
\tablecolumns{4}
\tablewidth{0pt}
\tablecaption{\textbf{\hst\ Data Used in this Study.}}
\tablehead{
\colhead{Dataset} &
\colhead{Date} & 
\colhead{Filter} &
\colhead{Exposure Time}
} 
\startdata
GO-9835   &  2003 Oct 30       &  F555W  &  48 $\times$ 155~s \\
GO-9835   &  2003 Oct 30       &  F814W  &  10 $\times$ 505~s \\
GO-9835   &  2003 Oct 30       &  F814W  &   5 $\times$ 25~s \\
GO-9835   &  2003 Oct 30       &  F814W  &   2 $\times$ 469~s \\
GO-10474  &  2006 Apr 04/05    &  F555W  &  48 $\times$ 155~s \\
GO-10474  &  2006 Apr 04/05    &  F814W  &   8 $\times$ 501~s \\
GO-10474  &  2006 Apr 04/05    &  F814W  &   4 $\times$  25~s \\
GO-10474  &  2006 Apr 04/05    &  F814W  &   4 $\times$ 508~s \\
GO-10350  &  2006 Apr 07       &  F555W  &   3 $\times$ 155~s \\
GO-10350  &  2006 Apr 07       &  F330W  &   2 $\times$ 1266~s \\
GO-10350  &  2006 Apr 07       &  F330W  &   4 $\times$ 1314~s \\
\enddata
\tablecomments{All images were captured with the High Resolution Channel on the Advanced Camera for Surveys. 
}\label{t:hst}
\end{deluxetable}

\begin{deluxetable}{l l l l l l } 
\tablewidth{0pt}

\tablecaption{{\textbf ST Photometry for Four Optical Counterparts}}

\tablehead{\colhead{star} & \colhead{RA (J2000)} & \colhead{Dec (J2000)} & \colhead{$U_{330}$ (2006)} & \colhead{$V_{555}$ (2003)} &  \colhead{$V_{555}$ (2006)} } 
\startdata

5	&	17$^h$36$^m$15.747$^s$&-44$^{\circ}$44'15.06'' &
21.99$\pm$0.02		&	22.48$\pm$0.02	&	23.15$\pm$0.02	\\ 
7	&	17$^h$36$^m$17.187$^s$&-44$^{\circ}$44'07.57'' & 
22.47$\pm$0.08		&	23.97$\pm$0.12		&	23.05	$\pm$0.31	\\ 
9	&	17$^h$36$^m$17.231$^s$&-44$^{\circ}$44'10.38'' &
21.63$\pm$0.02		&	19.95$\pm$0.01	&	22.23$\pm$0.06		\\ %
12	&	17$^h$36$^m$17.142$^s$&-44$^{\circ}$44'01.79'' &
22.35$\pm$0.04		&	22.07$\pm$0.02		&	22.84$\pm$0.08	  \\ 
\enddata

%\tablecomments{ }
\label{t:counter_phot}
\end{deluxetable}

%\begin{landscape}

\begin{deluxetable}{lccccccccr}
\rotate
\tablecolumns{10}
\tabletypesize{\footnotesize}
\tablewidth{7.5truein}
\tablecaption{\textbf{X-ray Sources in NGC 6388}}
\tablehead{
\multicolumn{2}{c}{\textbf{Source}} & \colhead{RA} & \colhead{Dec} & \colhead{Distance} & \multicolumn{2}{c}{Counts} & \multicolumn{2}{c}{$L_X$, ergs s$^{-1}$} & \colhead{Notes} \\
 (CX) & (CXOGLB J)  & (HH:MM:SS Err) & (DD:MM:SS Err) & (\arcsec)  & (0.5-1.5) & (1.5-6) & (0.5-6) & (0.5-2.5) & \\
}

\startdata
1  & 173617.6-444416 & 17:36:17.686  0.002 & -44:44:16.72  0.02 & 9.86 & $396.0^{+21.3}_{-19.9}$ & $129.5^{+13.3}_{-11.3}$ & $876.5^{+47.1}_{-41.2}$ & $842.9^{+43.9}_{-39.9}$ & q \\ 
2  & 173616.9-444409 & 17:36:16.937  0.002 & -44:44:09.85  0.02 & 3.42 & $268.5^{+17.8}_{-16.4}$ & $82.8^{+11.1}_{-9.0}$ & $663.7^{+43.0}_{-37.9}$ & $650.2^{+41.6}_{-37.4}$ & q \\ 
3  & 173617.3-444408 & 17:36:17.329  0.001 & -44:44:08.23  0.02 & 1.16 & $258.8^{+17.5}_{-16.1}$ & $68.9^{+10.5}_{-8.2}$ & $796.8^{+66.1}_{-51.2}$ & $700.6^{+47.4}_{-42.3}$ & q \\ 
4  & 173616.6-444423 & 17:36:16.633  0.003 & -44:44:23.48  0.03 & 16.50 & $34.6^{+7.4}_{-5.8}$ & $150.5^{+14.6}_{-12.2}$ & $660.0^{+69.6}_{-57.0}$ & $147.8^{+21.9}_{-17.2}$ & c* \\ 
5  & 173615.7-444415 & 17:36:15.751  0.003 & -44:44:15.02  0.03 & 17.06 & $79.1^{+10.4}_{-8.9}$ & $95.2^{+12.1}_{-9.7}$ & $444.4^{+52.4}_{-40.1}$ & $213.2^{+24.9}_{-20.4}$ & c \\ 
6  & 173617.3-444406 & 17:36:17.317  0.002 & -44:44:06.95  0.02 & 1.67 & $59.9^{+9.2}_{-7.7}$ & $104.9^{+12.6}_{-10.2}$ & $655.0^{+81.0}_{-60.5}$ & $315.3^{+37.9}_{-30.8}$ &  \\ 
7  & 173617.1-444407 & 17:36:17.195  0.003 & -44:44:07.59  0.03 & 0.68 & $45.8^{+8.3}_{-6.7}$ & $62.9^{+10.3}_{-7.8}$ & $328.6^{+54.1}_{-36.8}$ & $173.9^{+26.4}_{-20.5}$ & c \\ 
8  & 173618.1-444359 & 17:36:18.188  0.003 & -44:43:59.47  0.03 & 13.53 & $97.7^{+11.4}_{-9.8}$ & $10.9^{+5.9}_{-3.2}$ & $240.6^{+29.2}_{-24.5}$ & $240.6^{+29.2}_{-24.5}$ & q \\ 
9  & 173617.2-444410 & 17:36:17.245  0.003 & -44:44:10.31  0.04 & 2.10 & $44.5^{+8.2}_{-6.6}$ & $61.7^{+10.3}_{-7.7}$ & $270.8^{+46.7}_{-32.6}$ & $135.8^{+20.9}_{-16.0}$ & c \\ 
10  & 173617.5-444357 & 17:36:17.520  0.003 & -44:43:57.13  0.04 & 11.54 & $86.6^{+10.8}_{-9.3}$ & $13.8^{+6.1}_{-3.6}$ & $218.1^{+28.2}_{-23.1}$ & $214.5^{+26.8}_{-22.9}$ & q \\ 
11  & 173617.0-444403 & 17:36:17.017  0.003 & -44:44:03.100  0.04 & 5.62 & $41.7^{+8.0}_{-6.4}$ & $41.8^{+8.9}_{-6.3}$ & $269.7^{+54.8}_{-36.9}$ & $128.7^{+23.0}_{-17.1}$ & c? \\ 
12  & 173617.1-444401 & 17:36:17.158  0.004 & -44:44:01.76  0.04 & 6.48 & $33.7^{+7.3}_{-5.7}$ & $41.8^{+9.0}_{-6.3}$ & $264.9^{+54.1}_{-37.2}$ & $112.3^{+22.4}_{-16.5}$ & c \\ 
13  & 173617.0-444412 & 17:36:17.088  0.003 & -44:44:12.57  0.03 & 4.58 & $24.9^{+6.5}_{-4.9}$ & $27.9^{+7.8}_{-5.1}$ & $199.5^{+59.6}_{-34.6}$ & $109.5^{+25.5}_{-17.4}$ &  \\ 
14  & 173616.8-444412 & 17:36:16.883  0.004 & -44:44:12.78  0.05 & 5.81 & $24.8^{+6.5}_{-4.9}$ & $19.8^{+7.0}_{-4.3}$ & $125.1^{+40.1}_{-20.8}$ & $83.7^{+20.5}_{-14.3}$ &  \\ 
15  & 173617.3-444353 & 17:36:17.346  0.005 & -44:43:53.65  0.06 & 14.62 & $11.6^{+5.0}_{-3.3}$ & $26.7^{+7.7}_{-5.0}$ & $151.4^{+45.4}_{-29.2}$ & $53.7^{+17.0}_{-11.2}$ &  \\ 
16  & 173617.1-444413 & 17:36:17.158  0.004 & -44:44:13.07  0.04 & 4.93 & $14.9^{+5.5}_{-3.8}$ & $17.9^{+6.9}_{-4.0}$ & $144.0^{+53.4}_{-29.0}$ & $66.2^{+22.4}_{-14.3}$ &  \\ 
17  & 173618.0-444404 & 17:36:18.054  0.006 & -44:44:04.28  0.06 & 9.73 & $9.6^{+4.8}_{-3.0}$ & $21.8^{+7.2}_{-4.5}$ & $95.9^{+37.2}_{-19.7}$ & $53.5^{+17.0}_{-11.1}$ &  \\ 
18  & 173616.9-444411 & 17:36:16.969  0.005 & -44:44:11.79  0.05 & 4.43 & $9.8^{+4.8}_{-3.0}$ & $18.9^{+6.9}_{-4.1}$ & $100.1^{+42.0}_{-20.9}$ & $43.1^{+16.8}_{-9.8}$ &  \\ 
19  & 173617.4-444403 & 17:36:17.405  0.007 & -44:44:03.23  0.07 & 5.36 & $9.6^{+4.7}_{-3.0}$ & $14.8^{+6.5}_{-3.7}$ & $74.0^{+26.1}_{-15.6}$ & $30.9^{+14.6}_{-8.2}$ &  \\ 
20  & 173617.5-444406 & 17:36:17.558  0.007 & -44:44:06.64  0.08 & 3.94 & $11.6^{+5.0}_{-3.3}$ & $9.8^{+5.9}_{-2.9}$ & $62.1^{+32.9}_{-15.0}$ & $28.7^{+12.2}_{-7.5}$ &  \\ 
21  & 173617.8-444411 & 17:36:17.849  0.007 & -44:44:11.14  0.08 & 7.32 & $12.5^{+5.2}_{-3.4}$ & $7.8^{+5.5}_{-2.6}$ & $46.6^{+17.5}_{-10.6}$ & $29.7^{+13.1}_{-7.7}$ &  \\ 
22  & 173618.5-444415 & 17:36:18.555  0.008 & -44:44:15.57  0.08 & 16.05 & $11.6^{+5.0}_{-3.3}$ & $7.8^{+5.6}_{-2.6}$ & $39.3^{+19.2}_{-9.3}$ & $26.6^{+12.2}_{-6.5}$ &  \\ 
23  & 173617.3-444357 & 17:36:17.322  0.007 & -44:43:57.14  0.08 & 11.12 & $5.6^{+4.1}_{-2.2}$ & $13.7^{+6.4}_{-3.5}$ & $56.6^{+32.4}_{-14.3}$ & $23.9^{+12.6}_{-6.3}$ &  \\ 
24  & 173617.8-444344 & 17:36:17.863  0.008 & -44:43:44.67  0.08 & 24.52 & $7.6^{+4.4}_{-2.6}$ & $10.8^{+6.0}_{-3.1}$ & $60.5^{+32.2}_{-16.5}$ & $19.2^{+10.1}_{-5.7}$ &  \\ 
25  & 173613.9-444415 & 17:36:13.911  0.009 & -44:44:15.08  0.11 & 35.91 & $6.8^{+4.3}_{-2.5}$ & $7.8^{+5.6}_{-2.5}$ & $51.3^{+31.3}_{-15.0}$ & $18.3^{+11.8}_{-6.1}$ &  \\ 
26  & 173617.4-444410 & 17:36:17.440  0.009 & -44:44:10.17  0.10 & 3.06 & $8.5^{+4.6}_{-2.8}$ & $4.8^{+5.1}_{-2.0}$ & $27.1^{+13.8}_{-7.6}$ & $20.4^{+10.5}_{-6.2}$ &  \\ 
27  & 173616.9-444419 & 17:36:16.950  0.009 & -44:44:19.98  0.10 & 12.12 & $6.8^{+4.2}_{-2.5}$ & $5.8^{+5.3}_{-2.2}$ & $38.7^{+27.6}_{-12.3}$ & $11.0^{+7.8}_{-3.8}$ &  \\ 
28  & 173617.0-444359 & 17:36:17.083  0.008 & -44:43:59.86  0.09 & 8.53 & $8.8^{+4.5}_{-2.8}$ & $3.9^{+5.0}_{-1.7}$ & $35.6^{+32.6}_{-12.1}$ & $18.1^{+10.7}_{-5.3}$ &  \\ 
29  & 173617.1-444408 & 17:36:17.142  0.007 & -44:44:08.70  0.08 & 1.02 & $8.8^{+4.6}_{-2.8}$ & $2.9^{+4.8}_{-1.4}$ & $46.1^{+35.6}_{-14.6}$ & $27.5^{+14.2}_{-8.9}$ &  \\ 
30  & 173617.3-444438 & 17:36:17.300  0.010 & -44:44:38.00  0.11 & 29.81 & $1.9^{+3.2}_{-1.3}$ & $9.8^{+5.9}_{-2.9}$ & $38.5^{+28.0}_{-12.0}$ & $6.5^{+5.2}_{-2.7}$ &  \\ 
31  & 173618.2-444403 & 17:36:18.210  0.010 & -44:44:03.20  0.11 & 11.69 & $1.8^{+3.2}_{-1.2}$ & $8.8^{+5.7}_{-2.7}$ & $42.3^{+32.7}_{-15.1}$ & $13.2^{+9.5}_{-5.0}$ &  \\ 
32  & 173618.4-444408 & 17:36:18.465  0.011 & -44:44:08.38  0.12 & 13.28 & $2.7^{+3.4}_{-1.4}$ & $6.9^{+5.5}_{-2.4}$ & $41.5^{+33.8}_{-13.9}$ & $7.0^{+9.3}_{-3.3}$ &  \\ 
33  & 173617.2-444415 & 17:36:17.247  0.009 & -44:44:15.09  0.10 & 6.89 & $3.8^{+3.7}_{-1.8}$ & $4.9^{+5.1}_{-2.0}$ & $21.5^{+16.4}_{-6.9}$ & $17.4^{+13.3}_{-6.0}$ &  \\ 
34  & 173617.2-444416 & 17:36:17.235  0.009 & -44:44:16.86  0.10 & 8.66 & $3.8^{+3.7}_{-1.8}$ & $3.9^{+5.0}_{-1.6}$ & $23.5^{+22.1}_{-8.0}$ & $13.0^{+13.1}_{-4.9}$ &  \\ 
35  & 173617.3-444421 & 17:36:17.323  0.011 & -44:44:21.27  0.12 & 13.11 & $7.6^{+4.4}_{-2.6}$ & $0.0^{+4.2}_{-0.0}$ & $13.5^{+9.0}_{-5.0}$ & $13.5^{+9.0}_{-5.0}$ &  \\ 
36  & 173617.0-444351 & 17:36:17.091  0.012 & -44:43:51.02  0.13 & 17.24 & $6.6^{+4.3}_{-2.4}$ & $1.0^{+4.4}_{-0.8}$ & $18.6^{+12.7}_{-6.6}$ & $18.6^{+12.7}_{-6.6}$ &  \\ 
37  & 173617.6-444408 & 17:36:17.676  0.012 & -44:44:08.84  0.13 & 4.91 & $5.5^{+4.1}_{-2.2}$ & $1.9^{+4.6}_{-1.1}$ & $23.0^{+23.9}_{-9.4}$ & $13.9^{+9.8}_{-5.4}$ &  \\ 
38  & 173616.2-444415 & 17:36:16.233  0.012 & -44:44:15.43  0.14 & 12.75 & $3.6^{+3.6}_{-1.7}$ & $3.8^{+4.9}_{-1.9}$ & $10.9^{+8.4}_{-3.7}$ & $10.9^{+8.4}_{-3.7}$ &  \\ 
39  & 173617.6-444404 & 17:36:17.661  0.013 & -44:44:04.89  0.14 & 5.76 & $0.0^{+2.6}_{-0.0}$ & $6.8^{+5.5}_{-2.3}$ & $24.2^{+19.9}_{-9.0}$ & $9.5^{+9.6}_{-4.6}$ &  \\ 
40  & 173618.9-444358 & 17:36:18.959  0.013 & -44:43:58.08  0.14 & 21.13 & $3.9^{+3.7}_{-1.9}$ & $2.9^{+4.8}_{-1.4}$ & $20.8^{+26.5}_{-9.6}$ & $6.7^{+5.6}_{-2.8}$ &  \\ 
41  & 173617.0-444405 & 17:36:17.069  0.010 & -44:44:05.94  0.11 & 2.78 & $6.7^{+4.3}_{-2.5}$ & $0.0^{+4.2}_{-0.0}$ & $16.5^{+10.9}_{-6.4}$ & $16.5^{+10.9}_{-6.4}$ &  \\ 
42  & 173616.8-444406 & 17:36:16.842  0.010 & -44:44:06.59  0.12 & 4.33 & $6.7^{+4.3}_{-2.4}$ & $0.0^{+4.2}_{-0.0}$ & $17.5^{+11.0}_{-6.7}$ & $17.5^{+11.0}_{-6.7}$ &  \\ 
43  & 173619.0-444341 & 17:36:19.002  0.013 & -44:43:41.79  0.13 & 32.55 & $3.7^{+3.7}_{-1.8}$ & $2.8^{+4.8}_{-1.4}$ & $19.8^{+24.8}_{-8.8}$ & $7.2^{+8.0}_{-3.0}$ &  \\ 
44  & 173618.5-444412 & 17:36:18.521  0.013 & -44:44:12.24  0.14 & 14.45 & $2.6^{+3.4}_{-1.4}$ & $3.9^{+5.0}_{-1.7}$ & $13.1^{+12.8}_{-4.7}$ & $9.8^{+10.1}_{-3.7}$ &  \\ 
45  & 173617.5-444402 & 17:36:17.592  0.014 & -44:44:02.16  0.15 & 7.39 & $3.6^{+3.7}_{-1.7}$ & $1.9^{+4.6}_{-1.1}$ & $14.6^{+14.2}_{-5.6}$ & $10.9^{+11.2}_{-4.7}$ &  \\ 
46  & 173617.5-444413 & 17:36:17.550  0.014 & -44:44:13.88  0.15 & 6.53 & $1.8^{+3.2}_{-1.2}$ & $3.8^{+5.0}_{-1.6}$ & $23.1^{+26.8}_{-9.6}$ & $6.1^{+7.9}_{-3.5}$ &  \\ 
47  & 173616.5-444410 & 17:36:16.593  0.013 & -44:44:10.63  0.14 & 7.09 & $4.6^{+3.9}_{-2.0}$ & $0.9^{+4.4}_{-0.8}$ & $12.7^{+13.5}_{-5.3}$ & $8.4^{+8.0}_{-3.8}$ &  \\ 
48  & 173614.7-444430 & 17:36:14.769  0.016 & -44:44:30.79  0.18 & 34.53 & $0.9^{+3.0}_{-0.8}$ & $3.8^{+5.0}_{-1.7}$ & $15.6^{+23.2}_{-8.2}$ & $6.5^{+7.9}_{-3.0}$ &  \\ 
49  & 173618.1-444419 & 17:36:18.152  0.015 & -44:44:19.24  0.17 & 14.85 & $1.8^{+3.2}_{-1.2}$ & $2.9^{+4.8}_{-1.4}$ & $11.5^{+14.3}_{-5.1}$ & $3.5^{+4.5}_{-1.8}$ &  \\ 
50  & 173616.0-444408 & 17:36:16.036  0.015 & -44:44:08.14  0.17 & 12.61 & $1.8^{+3.2}_{-1.2}$ & $2.9^{+4.8}_{-1.5}$ & $16.9^{+23.2}_{-8.6}$ & $7.7^{+8.1}_{-3.8}$ &  \\ 
51  & 173616.6-444421 & 17:36:16.689  0.013 & -44:44:21.94  0.15 & 14.85 & $0.9^{+3.0}_{-0.8}$ & $2.9^{+4.7}_{-1.6}$ & $8.7^{+9.8}_{-4.0}$ & $8.7^{+9.8}_{-4.0}$ &  \\ 
52  & 173618.0-444412 & 17:36:18.026  0.015 & -44:44:12.70  0.16 & 9.70 & $2.9^{+3.5}_{-1.6}$ & $1.0^{+4.4}_{-0.8}$ & $5.4^{+5.4}_{-2.5}$ & $5.4^{+5.4}_{-2.5}$ &  \\ 
53  & 173617.3-444359 & 17:36:17.363  0.015 & -44:43:59.24  0.16 & 9.10 & $2.9^{+3.5}_{-1.6}$ & $0.9^{+4.4}_{-0.8}$ & $5.9^{+5.9}_{-2.7}$ & $5.9^{+5.9}_{-2.7}$ &  \\ 
54  & 173616.0-444421 & 17:36:16.079  0.017 & -44:44:21.33  0.19 & 17.88 & $0.9^{+2.9}_{-0.8}$ & $2.8^{+4.8}_{-1.4}$ & $14.8^{+23.7}_{-8.3}$ & $2.3^{+4.3}_{-1.4}$ &  \\ 
55  & 173618.3-444434 & 17:36:18.308  0.018 & -44:44:34.22  0.19 & 28.48 & $1.7^{+3.2}_{-1.1}$ & $1.9^{+4.6}_{-1.1}$ & $9.3^{+13.4}_{-4.6}$ & $4.7^{+7.6}_{-2.5}$ &  \\ 
56  & 173617.0-444358 & 17:36:17.074  0.015 & -44:43:58.25  0.16 & 10.08 & $0.9^{+2.9}_{-0.8}$ & $1.9^{+4.6}_{-1.1}$ & $9.2^{+13.5}_{-4.7}$ & $9.2^{+13.5}_{-4.7}$ &  \\ 
57  & 173618.0-444332 & 17:36:18.099  0.019 & -44:43:32.12  0.21 & 37.28 & $0.8^{+2.9}_{-0.7}$ & $2.0^{+4.6}_{-1.3}$ & $9.2^{+11.1}_{-4.9}$ & $2.4^{+6.3}_{-2.1}$ &  \\ 
58  & 173617.2-444341 & 17:36:17.246  0.019 & -44:43:41.74  0.21 & 26.47 & $2.7^{+3.4}_{-1.4}$ & $0.0^{+4.2}_{-0.0}$ & $4.5^{+7.0}_{-2.5}$ & $4.5^{+7.0}_{-2.5}$ &  \\ 
59  & 173616.7-444400 & 17:36:16.752  0.019 & -44:44:00.39  0.22 & 9.27 & $2.5^{+3.4}_{-1.4}$ & $0.0^{+4.2}_{-0.0}$ & $6.5^{+8.6}_{-3.8}$ & $6.5^{+8.6}_{-3.8}$ &  \\ 
60  & 173616.5-444416 & 17:36:16.579  0.020 & -44:44:16.52  0.22 & 10.75 & $0.7^{+2.9}_{-0.7}$ & $1.8^{+4.6}_{-1.1}$ & $7.6^{+12.9}_{-4.3}$ & $3.2^{+6.9}_{-2.1}$ &  \\ 
61  & 173617.0-444329 & 17:36:17.088  0.019 & -44:43:29.88  0.21 & 38.36 & $0.9^{+2.9}_{-0.7}$ & $1.0^{+4.4}_{-0.8}$ & $11.7^{+23.0}_{-8.1}$ & $2.4^{+6.3}_{-2.1}$ &  \\ 
\enddata
\tablecomments{
Names, positions, distance from center of NGC 6388, counts in two X-ray energy bands (energies given in keV), and estimated X-ray luminosities (in units of $10^{30}$ ergs s$^{-1}$) of X-ray sources associated with NGC 6388.  
The errors after the position represent 
the $1\sigma$ uncertainties in the relative positions of the sources,
derived from ACIS\_EXTRACT centroiding.  The counts in each band are the
numbers of photons within the source regions of Figure
1. Luminosities are computed from the corrected photon fluxes in
several narrow bands, see text. Notes indicate short-term variability 
(V = 99\% confidence, V? = 95\% confidence),
years-timescale variability between 2000 and 2003 (Y), and possible
identifications (q = qLMXB, q? = qLMXB 
candidate, c = CV with optical counterpart identified, c* = CV identified from X-ray properties, c? = possible optical counterpart identification). 
}\label{t:xraysources}
\end{deluxetable}

\clearpage

%Need to rotate this.
\begin{deluxetable}{cccccccccccccc}
\tabletypesize{\scriptsize}
\rotate
\tablewidth{7.5truein}
\tablecaption{\textbf{Spectral Fits to Bright NGC 6388 Sources}}
\tablehead{
\colhead{\textbf{CX}}  & \multicolumn{5}{c}{H-atmosphere + Power-law} & \multicolumn{4}{c}{MEKAL} & \multicolumn{4}{c}{Power-law} \\
 & \colhead{kT} & \colhead{$N_H$} & \colhead{PL} & \colhead{Good-} & \colhead{$L_X$} & \colhead{kT}  & \colhead{$N_H$} & \colhead{Good-} & \colhead{$L_X$} &  \colhead{$\alpha$} & \colhead{$N_H$} & \colhead{Good-} & \colhead{$L_X$} \\
 & & & fraction & ness & & & & ness & & & & ness & \\
 & (eV) & \tiny{($\times10^{21}$} & (\%) & (\%) & \tiny{($\times10^{32}$}& (keV) &  \tiny{($\times10^{21}$}  & (\%) & \tiny{($\times10^{32}$} & &  \tiny{($\times10^{21}$} & (\%) & \tiny{($\times10^{32}$} \\ 
 & & \tiny{ cm$^{-2}$)} & & & \tiny{ergs s$^{-1}$)} & &  \tiny{cm$^{-2}$)} & & \tiny{ergs s$^{-1}$)} & &  \tiny{cm$^{-2}$)} & & \tiny{ergs s$^{-1}$)} 
}
\startdata  %$^{+}_{-}$
\hline \\
1 & 130$^{+2}_{-2}$ & 4.0$^{+0.4}_{-0.4}$ & $0^{+4}_{-0}$ & 51 & $15^{+1}_{-1}$ & 0.49 & 8.9 & 100 & 62 & $5.4^{+0.6}_{-0.5}$ & $8.6^{+0.1}_{-0.1}$ & 8 & 130$^{+110}_{-50}$ \\
2 & 119$^{+3}_{-3}$ & 3.5$^{+0.4}_{-0.4}$ & $0^{+7}_{-0}$  & 10 & 9.9$^{+1.2}_{-1.2}$ & 1.3$^{+0.1}_{-0.2}$ & 2.2$^{+0.2}_{-0}$ & 13.4 & $6.7^{+0.6}_{-0.6}$ & 5.0$^{+0.6}_{-0.6}$ & 6.8$^{+1.3}_{-1.2}$ & 2.5 & $49^{+46}_{-21}$ \\
3  & $117^{+4}_{-4}$ & $3.2^{+0.5}_{-0.5}$ & $18^{+11}_{-7}$ & 40 & $11.0^{+2.3}_{-2.0}$ & 1.1 & 2.2 & 95 & $7.3$ & $4.5^{+0.7}_{-0.6}$ & $6.0^{+1.4}_{-1.3}$ & 79 & $37^{+36}_{-15}$ \\
8 & $92^{+4}_{-4}$ & $2.5^{+0.5}_{-0.3}$ & $0^{+8}_{-0}$ & 72 & $3.0^{+0.6}_{-0.6}$ &
 $0.8^{+0.1}_{-0.1}$ & $2.2^{+0.4}_{-0}$ & 89 & $2.3^{+0.4}_{-0.3}$ &
$6.6^{+1.9}_{-1.5}$ & $0.8^{+0.3}_{-0.3}$ & 24 & $5^{+29}_{-4}$ \\
10 & $91^{+5}_{-4}$ & $2.7^{+0.7}_{-0.5}$ & $0^{+13}_{-0}$ & 6 & $2.8^{+0.7}_{-0.6}$ &
$0.5^{+0.2}_{-0.3}$ & $3.9^{+5.0}_{-1.7}$ & 85 & $4.0^{+96}_{-2.0}$ &
$6.2^{+1.9}_{-1.4}$ & $7.4^{+3.2}_{-2.5}$ & 9 & $30^{+170}_{-20}$ \\
\hline
\hline
11 & $87^{+11}_{-20}$ & $5.5^{+2.0}_{-2.1}$ & $60^{+28}_{-23}$ & 21 & $5.5^{+0.7}_{-0.5}$ &
 $5.0^{+14}_{-2.3}$ & $2.7^{+1.6}_{-0.4}$ & 6 & $3.2^{+1.0}_{-0.7}$ &
$2.0^{+0.6}_{-0.6}$ & $3.8^{+2.2}_{-1.6}$ & 14 & $3.8^{+1.2}_{-0.8}$ \\
5 & $49^{+31}_{-49}$ & $2.2^{+1.8}_{-0}$ & $98^{+2}_{-21}$ & 72 & $6^{+1}_{-1}$ &
$8.6^{+15}_{-3.8}$ & $2.2^{+.6}_{-0}$ & 61 & $5.5^{+1.2}_{-0.9}$ &
$1.6^{+.4}_{-.2}$ & $2.2^{+1.3}_{-0}$ & 59 & $5.9^{+1.2}_{-0.9}$ \\
4 & $67^{+24}_{-67}$ & $10.2^{+3.3}_{-2.5}$ & $95^{+5}_{-14}$ & 41 & $13^{+4}_{-3}$ &
80 & 7.6 & 99 & $11.0$ &
$0.9^{+0.4}_{-0.4}$ & $5.0^{+3.4}_{-2.5}$ & 26 & $12^{+3}_{-2}$ \\
7 & $<96$ & $2.5^{+4.3}_{-0.3}$ & $100^{+0}_{-45}$ & 3 & $5.1^{+2.6}_{-0.9}$ &
$4.4^{+8.1}_{-1.8}$ & $3.1^{+1.6}_{-0.9}$ & 23 & $4.2^{+1.1}_{-0.8}$ &
$2.0^{+0.6}_{-0.5}$ & $4.2^{+}_{-}$ & 24 & $5.1^{+1.6}_{-1.0}$ \\
9 & $89^{+13}_{-24}$ & $6.7^{+2.4}_{-1.6}$ & $61^{+30}_{-33}$ & 8 & $6.4^{+3.6}_{-2.0}$ &
$4.2^{+6.4}_{-1.7}$ & $4.0^{+1.8}_{-1.5}$ & 63 & $3.7^{+0.8}_{-0.8}$ &
$2.2^{+0.6}_{-0.5}$ & $5.4^{+2.5}_{-2.1}$ & 29 & $4.6^{+1.8}_{-1.0}$ \\
6 & $0^{+87}_{-0}$ & $2.8^{+0.9}_{-0.3}$ & $100^{+0}_{-22}$ & 31 & $9.9^{+2.7}_{-1.5}$ &
$9.0^{+37}_{-4.7}$ & $2.7^{+1.2}_{-0.5}$ & 40 & $8.9^{+1.8}_{-0.9}$  & 
$1.6^{+0.4}_{-0.4}$ & $3.2^{+1.6}_{-1.0}$ & 18 & $9.7^{+1.8}_{-1.6}$ \\
\enddata
\tablecomments{Spectral fits to sources with more than 80 counts.  Errors are 90\% confidence for one parameter of interest.
The hydrogen column density ($N_H$) to the cluster is $2.2\times10^{21}$ cm$^{-2}$, which was used as a lower bound.  
$L_X$ in erg s$^{-1}$ for 0.5-10 keV, assuming distance of 10 kpc.  Hydrogen atmosphere NS fits assume R=10 km, M=1.4 \msun NSs, while the power-law photon index was fixed to 1.5.  ``Goodness'' is the fraction of simulations with a smaller C-statistic than the given spectral fit; large ``goodness'' values indicate a poor fit, for which no errors are provided.
\label{t:spectra}}
\end{deluxetable}

\clearpage

\begin{deluxetable}{r@{ }l l  }

\tablecaption{Population distribution comparisons.}
\tablewidth{0pc}
\tablehead {\multicolumn{2}{c}{compared pops} & \colhead{Prob.}} 
\startdata
\sidehead{Optical:}
RHB vs.& 20 w/o & 0.32 \\
BS vs.& 20 w/o & 0.004 \\
RHB vs.& BHB & 0.67  \\ 
RHB vs.& EBHB & 0.96 \\ 
BHB vs.& EBHB & 0.91 \\ 
\tableline
\sidehead{X-ray:}
non-qLMXB vs.& qLMXB & 0.48 \\
bright vs.& faint & 0.09 \\

\enddata

\tablecomments{
The probability that the listed populations are drawn from the same underlying distribution as determined by a K-S test.
`20' signifies all stars with $V_{555}<20.$ `w/o'
specifies that the RHB or BS were removed from the `20' sample before comparing
to their respective groups. 
} \label{t:pops} 
\end{deluxetable}

\clearpage

\begin{deluxetable}{l l l l }
\tablewidth{0pc}
\tablecaption{Generalized King Model Fit Parameters}
\tablehead{\colhead{pops} & \colhead{q} & \colhead{$r_c$} & \colhead{$\alpha$} } 
\startdata

BS        & 1.80 $\pm$ 0.27  & 4.39  $\pm$ 0.48     &  -4.39 $\pm$ 0.27 \\
RHB       &  0.97 $\pm$ 0.10 & 7.41 $\pm$ 0.93 &  -1.92 $\pm$ 0.33 \\
all X-ray 
with bkd corr & 1.51 $\pm$ 0.12 & 4.99 $\pm$ 0.37 &  -3.54 $\pm$ 0.37 \\
qLMXBs     & 1.68 $\pm$ 0.31 & 4.62 $\pm$ 0.65 &  -4.03 $\pm$ 0.92 \\
non-qLMXBs     & 1.40 $\pm$ 0.10 & 5.33 $\pm$ 0.32 &  -3.20 $\pm$ 0.31 \\
CVs       & 1.57 $\pm$ 0.48 & 4.85 $\pm$ 1.07 &  -3.71 $\pm$ 1.50 \\
Bright X-ray ($>$40 cnts) & 1.82 $\pm$ 0.34 & 4.35 $\pm$ 0.55 &  -4.45 $\pm$ 1.00 \\
Faint X-ray & 1.31 $\pm$ 0.11 & 5.59 $\pm$ 0.39 &  -2.94 $\pm$ 0.33 \\

\enddata
\tablecomments{Best-fit generalized King Model parameters for several different populations. Errors are derived from bootstrap statistics.}
\label{t:pops2}
\end{deluxetable}

\clearpage

\begin{deluxetable}{lcccccl}
\tabletypesize{\normalsize}
\tablecolumns{7}
\tablewidth{0pt}
\tablecaption{Approximate Collision Parameters for a Sample of Clusters}
\tablehead{
\colhead{Cluster} &
\multicolumn{2}{c}{$\Gamma$} & 
\multicolumn{2}{c}{Sources Detected} & 
\colhead{Background} &
\colhead{References}\\
\colhead{(1)} & (2) & (3) & (4) & (5)& (6) & \colhead{(7)}  
% & & (1) & (1) & {\footnotesize Updated} & &  \colhead{{\footnotesize for updated}} \\
% & & & & & & \colhead{{\footnotesize detections} }
% & {\footnotesize Approximated} & {\footnotesize \citet{Pooley03}}  & {\footnotesize \citet{Pooley03}} & {\footnotesize Recent Values} & &
} 

\startdata

NGC 6266     & 516 & 500     & 51 & \nodata     & 2-3 &\\ 
47 Tuc       & 235 & 434     & 45 & 79  	& $\sim16$  &  \citet{Heinke05} \\ 
NGC 6626     & 148 & 186     & 26 & \nodata 	& 2-3 &\\ 
NGC 6093     & 109 & 166     & 17 & 19  	& 1-2 & \citet{Heinke03} \\ 
NGC 5904     & 40  & 69      & 16 & \nodata	& 4-7 &\\ 
$\omega$ Cen & 23  & 49      & 28 & 60  	& 19-24 &\citet{Haggard09} \\ 
NGC 6752     & 81  & 38      & 11 & \nodata	& 2-3 & \\ 
NGC 7099     & 23  & 18      & 7  & 11  	& 1-2 & \citet{Lugger07}\\ 
NGC 6121     & 8   & 13      & 6  & 6   	& 1-3 & \citet{Bassa04} \\ 
NGC 6397     & 12  & 5.9     & 12 & 11  	& 0-1 & \citet{Bogdanov10}\\
NGC 6366     & 2.1 & 2.3     & 4  & 5   	& 2-4 & \citet{Bassa08}\\ 
NGC 6388     & 515 & \nodata & \nodata & 61 	& 1-2 & 
\enddata
\tablecomments{Col. (2): The approximate values of $\Gamma$ are shown against col. (3): the value of $\Gamma$ reported by \citet{Pooley03}.   Col. (4): Numbers of detected sources as reported by \citet{Pooley03} are compared to col. (5): updated values from the literature for a limit of $4\times10^{30}$\ergs. The limit for NGC 6388 is $\sim5\times10^{30}$\ergs. Col. (6): predicted background counts. Col. (7): sources for the updated results in col. (5).
}
\label{t:collision}
%\tablerefs{(1) \citet{Heinke05} 
%(2)\citet{Heinke03}
%(3)\citet{Haggard09} 
%(4)\citet{Lugger07} 
%(5)\citet{Bassa04} 
%(6)\citet{Bogdanov10}} 
\end{deluxetable}
%\tablenotetext{a}{\citet{Heinke05}} 
%\tablenotetext{b}{\citet{Heinke03}, $L_x > 1 \times 10^{31} \ergs$} 
%\tablenotetext{c}{\citet{Haggard09}} 
%\tablenotetext{d}{\citet{Lugger07}} 
%\tablenotetext{e}{\citet{Bassa04}} 
%\tablenotetext{f}{\citet{Bogdanov10}} 
%\tablenotetext{g}{\citet{Bass08}} 

\end{document}